\ifx\pdfsuppresswarningpagegroup\undefined\else\pdfsuppresswarningpagegroup=1\fi\relax
\RequirePackage{silence}
\PassOptionsToPackage{unicode, pdfprintscaling=None, colorlinks}{hyperref}

\documentclass[aps,prc,reprint,nofootinbib,preprintnumbers,floatfix,unsortedaddress,superscriptaddress,longbibliography]{revtex4-2}

\usepackage{orcidlink,multirow,latexsym,amsmath,amssymb,amsfonts,bm,bbm,microtype,xspace,placeins,float,mathtools,xcolor,makecell,tabularx}

\usepackage[normalem]{ulem}
\usepackage[export]{adjustbox}
\usepackage{hyperref}
\usepackage[capitalise]{cleveref}

\usetikzlibrary{decorations.markings, arrows, decorations.pathmorphing}
\graphicspath{{pic/}}
\allowdisplaybreaks

\newlength{\FigureWidth}
\newcolumntype{L}[1]{>{\raggedright\arraybackslash}p{#1}}
\newcolumntype{C}[1]{>{\centering\arraybackslash}p{#1}}
\newcolumntype{R}[1]{>{\raggedleft\arraybackslash}p{#1}}
\newcommand\TopRule{\Xhline{0.08em}}

\newcommand\MidRule{\Xhline{0.03em}}
\newcommand\BotRule{\Xhline{0.08em}}

\makeatletter
\newcommand\showtitleinbib{{\escapechar=`\\ \immediate\write\@auxout{%
\csname citation{REVTEX42Control}\endcsname^^J%
\csname citation{apsrev42Control}\endcsname
}}}
\makeatother

\newcommand{\bfe}{{\mathbf{e}}}

\newcommand{\bfn}{{\mathbf{n}}}

\newcommand{\ba}{{\pmb{\alpha}}}

\newcommand{\lml}{{\lambda_\ell}}

\newcommand{\cH}{{\cal H}}
\newcommand{\chU}{{\hat{\cal U}}}
\newcommand{\chUp}{{\hat{\cal U}_P}}
\newcommand{\chE}{{\hat{\cal E}}}

\newcommand{\phys}{{\mathrm{phys}}}

\newcommand{\Tr}{{\mathrm{Tr}}}
\newcommand{\mdck}{{\{\lambda_s\},\{\lambda_\ell\},\{\alpha_s\}}}

\newcommand{\mdc}{{[\{\lambda_s\},\{\lambda_\ell\}]}}
\newcommand{\dck}{{\{\lambda_\ell\},\{\alpha_s\}}}
\newcommand{\dckp}{{\{\lambda_\ell'\},\{\alpha_s'\}}}
\newcommand{\dc}{{\{\lambda_\ell\}}}
\newcommand{\dcxy}{{\{\lambda_\ell\};\allowbreak\lambda_x,\lambda_y}}
\newcommand{\dcxyk}{{\{\lambda_\ell\},\{\alpha_s\};\allowbreak\lambda_x,\lambda_y}}

\newcommand{\ldim}{{{\cal D}(\cH_s^g)}}
\newcommand{\Hsg}{{\cH_s^g}}
\newcommand{\ungroup}[1]{#1}
\newcommand{\withbreak}[1]{\expandafter\ungroup#1}

\def\dounbracket[#1]{#1}
\newcommand{\txtw}{{\mathsf{w}}}

\newcommand{\SU}{{\mathrm{SU}}}
\def\bra#1{\left\langle #1\right|}
\def\ket#1{\left| #1\right\rangle}

\newcommand\dm{{d}}

\renewcommand\vec\mathbf

\newcommand\redsout{\bgroup\markoverwith{\textcolor{red}{\rule[0.5ex]{2pt}{1.4pt}}}\ULon}

\newcommand\dootimesall[2]{\ifx0#1\else\mathbf{#1}\ifx0#2\else\def\mytmp{\otimes\dootimesall{#2}}\expandafter\expandafter\expandafter\mytmp\fi\fi}

\newcommand\otimesall[1]{\dootimesall#10}

\newcommand\bket[2]{\ket{\mathbf#1\mathbf#2}}
\newcommand\bbra[2]{\bra{\mathbf#1\mathbf#2}}
\newcommand\bop[4]{\bket#1#2\bbra#3#4}

\usepackage{relsize}

\usepackage[acronyms, nohypertypes={acronym}, nopostdot, style=super, nonumberlist, toc]{glossaries}
\setacronymstyle{long-sc-short}
\glsenableentrycount

\newcommand\ac[1]{\gls{#1}}

\newcommand\acp[1]{\glspl{#1}}

\newacronym{PNA}{pna}{particle-number-algorithm}
\newacronym{SFA}{sfa}{spin-flip-algorithm}

\newacronym{WF}{wf}{Wilson-Fisher}
\newacronym{AF}{af}{asymptotically free}

\newacronym{RG}{rg}{renormalization group}

\newacronym{QIS}{qis}{Quantum Information Science}
\newacronym{PPT}{ppt}{positive-semidefinite partial transpose}
\newacronym{KS}{ks}{Kogut-Susskind}
\newacronym{NPT}{npt}{negative partial transpose}

\newacronym{AS}{as}{Anti-Symmetric}

\newacronym[longplural={conformal field theories}]{CFT}{cft}{conformal field theory}
\newacronym[longplural={lattice field theories}]{LFT}{lft}{lattice field theory}
\newacronym[longplural={effective field theories}]{EFT}{eft}{effective field theory}
\newacronym[longplural={quantum field theories}]{QFT}{qft}{quantum field theory}
\newacronym[longplural={lattice gauge theories}]{LGT}{lgt}{lattice gauge theory}
\newacronym[longplural={monomer-dimer tensor-networks}]{MDTN}{mdtn}{monomer-dimer tensor-network}

\newacronym[]{DMRG}{dmrg}{Density Matrix Renormalization Group}
\newacronym[]{TFIM}{tfim}{Transverse Field Ising Model}

\newacronym[]{LOCC}{locc}{Local Operations and Classical Communicaton}
\newacronym[]{OBC}{obc}{open boundary conditions}

\newacronym{MPS}{mps}{matrix product states}

\newacronym{JLP}{jlp}{Jordan-Lee-Preskill}

\newacronym{BBN}{bbn}{big bang nucleosynthesis}

\newacronym{LEC}{lec}{low-energy constant}

\newacronym{QCD}{qcd}{quantum chromodynamics}
\newacronym{MC}{mc}{Monte Carlo}

\newacronym{IR}{ir}{infrared}
\newacronym{UV}{uv}{ultraviolet}

\newacronym{QED}{qed}{quantum electrodynamics}
\newacronym{SNR}{snr}{signal-to-noise ratio}

\newacronym{NLSM}{nlsm}{nonlinear sigma model}

\newacronym{CL}{cl}{Complex Langevin}

\newacronym{CSA}{csa}{Cartan subalgebra}

\newacronym{SSB}{ssb}{spontaneous symmetry breaking}

\newacronym{AFQMC}{afqmc}{auxiliary field quantum Monte Carlo}
\newacronym{iHMC}{ihmc}{imaginary-mass Hybrid Monte Carlo}

\newacronym{MCMC}{mcmc}{Markov Chain Monte Carlo}

\newacronym{QI}{qi}{quantum information}

\newacronym{irrep}{{\rm irrep}}{irreducible representation}

\newacronym{ASQR}{asqr}{antisymmetric qubit regularization}

\newif\ifshowchanges\showchangesfalse
\ifshowchanges
\usepackage{soul}

\newcommand\tanmoy[2][]{{\setstcolor{red}\setul{1ex}{0.25ex}%
\ifx\undefined#1\undefined\else\st{#1}\fi\color{red}{#2}}}
\newcommand\shailesh[2][]{{\setstcolor{magenta}\setul{1ex}{0.25ex}%
\ifx\undefined#1\undefined\else\st{#1}\fi\color{magenta}{#2}}}
\newcommand\siew[2][]{{\setstcolor{cyan}\setul{1ex}{0.25ex}%
\ifx\undefined#1\undefined\else\st{#1}\fi\color{cyan}{#2}}}
\else
\newcommand\tanmoy[2][]{#2}
\newcommand\shailesh[2][]{#2}
\newcommand\siew[2][]{#2}
\fi

\begin{document}

\title{Monomer-dimer tensor-network basis for qubit-regularized lattice gauge theories}
\author{Shailesh Chandrasekharan\orcidlink{0000-0002-3711-4998}}
\email{sch27@duke.edu}
\affiliation{ Department of Physics, Box 90305, Duke University, Durham, North Carolina 27708, USA}
\author{Rui Xian Siew\orcidlink{0000-0002-0745-8853}}%
 \email{ruixian.siew@duke.edu}
\affiliation{ Department of Physics, Box 90305, Duke University, Durham, North Carolina 27708, USA}
\author{Tanmoy Bhattacharya\,\orcidlink{0000-0002-1060-652X}}
\email{tanmoy@lanl.gov}
\affiliation{Theoretical Division, Los Alamos National Laboratory, Los Alamos, New Mexico 87545, USA}

\date{\today}

\begin{abstract}  
Traditional $\mathrm{SU}(N)$ \glspl{LGT} can be formulated using an orthonormal basis constructed from the \glspl{irrep} $V_{\lambda}$ of the $\mathrm{SU}(N)$ gauge symmetry. On a lattice, the elements of this basis are tensor networks comprising dimer tensors on the links labeled by a set of \glspl{irrep} $\{\lambda_\ell\}$ and monomer tensors on sites labeled by $\{\lambda_s\}$. These tensors naturally define a local site Hilbert space, $\mathcal{H}^g_s$, on which gauge transformations act. Gauss's law introduces an additional index $\alpha_s = 1, 2, \dots, {\cal D}(\cH_s^g)$ that labels an orthonormal basis of the gauge-invariant subspace of $\mathcal{H}^g_s$.  
This \gls{MDTN} basis, $\ket{\{\lambda_s\},\{\lambda_\ell\},\{\alpha_s\}}$, of the physical Hilbert space enables the construction of new qubit-regularized $\mathrm{SU}(N)$ gauge theories that are free of sign problems while preserving key features of traditional \glspl{LGT}. Here, we investigate finite-temperature confinement-deconfinement transitions in a simple qubit-regularized $\mathrm{SU}(2)$ and $\mathrm{SU}(3)$ gauge theory in $d=2$ and $d=3$ spatial dimensions, formulated using the \gls{MDTN} basis, and show that they reproduce the universal results of traditional \glspl{LGT} at these transitions.  Additionally, in $d=1$, we demonstrate using a plaquette chain that the string tension at zero temperature can be continuously tuned to zero by adjusting a model parameter that plays the role of the gauge coupling in traditional \glspl{LGT}.  
\end{abstract}

\preprint{LA-UR-24-32125}

\maketitle


\section{Introduction}
\label{sec-1}

Quantum field theories are a special class of quantum systems defined on infinite-dimensional Hilbert spaces. Qubit regularization of a \ac{QFT} is the idea of formulating these theories through a limiting process, starting from quantum mechanical systems defined on finite-dimensional Hilbert spaces \cite{Chandrasekharan:2025pil}. Such a regularization, particularly in a Hamiltonian formulation, is well-suited for studying \acp{QFT} using quantum technologies \cite{Banuls:2019bmf,Bauer:2023qgm,DiMeglio:2023nsa}. 

While the Hamiltonian lattice regularization of a \ac{QFT} with infinite-dimensional local Hilbert spaces is well known \cite{Kogut:1974ag}, the idea of starting with a finite-dimensional Hilbert space for both spin and gauge theories was originally proposed in the D-theory approach \cite{Chandrasekharan:1996ih,Chandrasekharan:1998ck,Brower:1997ha,Brower:2003vy} and has recently gained popularity \cite{Banuls:2017ena,Alexandru:2019ozf,Alexandru:2019nsa,PhysRevD.109.094502}. Most studies assume that in the limiting process, the local Hilbert space on each lattice site will need to be extended indefinitely to formulate asymptotically free \acp{QFT}, such as Yang-Mills theories and \ac{QCD}. For this reason, in the D-theory approach, an extra dimension (or equivalently, a flavor index) was introduced at each spatial lattice site, allowing for a systematic increase in the local Hilbert space.

This common belief that recovering asymptotic freedom ultimately requires increasing the local lattice Hilbert space dimension to infinity was challenged by the discovery of two examples, where asymptotic freedom emerged in qubit-regularized $\dm=1$ $O(3)$ and $O(2)$ spin-models with just a four-dimensional local Hilbert space \cite{Bhattacharya:2020gpm,Maiti:2023kpn}. Such a possibility had been suggested within the D-theory approach \cite{Chandrasekharan:1998ck}, but a concrete realization had not been found until these examples were discovered. From a \ac{RG} perspective, this is not particularly surprising, since \acp{QFT} emerge near fixed points of \ac{RG} flows, which can be accessed by tuning the couplings of lattice models to appropriate quantum critical points. If only a few relevant parameters exist at the fixed point, one should be able to tune a small number of lattice parameters non-perturbatively to reach the critical point. In this approach, the finiteness of the local lattice Hilbert space does not preclude an infinite-dimensional Hilbert space in the \hbox{\ac{QFT}} because the length scales of the emerging \hbox{\ac{QFT}} are scaled by the diverging correlation length of the lattice theory.

The \acp{QFT} studied in \cite{Bhattacharya:2020gpm,Maiti:2023kpn} had only one relevant parameter and, surprisingly, could be reached even without fine-tuning. However, the \ac{RG} flow was markedly different from traditional expectations. At the quantum critical point, the infrared physics was governed by a completely different conformal field theory. However, when a small relevant perturbation was introduced, the desired asymptotically free quantum field theory emerged as a crossover phenomenon. These discoveries are exciting, as they suggest that qubit regularization has the potential to reveal new types of \ac{RG} flows through which \acp{QFT} can emerge, particularly for gauge theories.

The motivation behind our current work is to establish a foundation for a more systematic exploration of qubit-regularized $\SU(N)$ gauge theories, starting with the Hilbert space of traditional lattice gauge theories in a basis that is well-suited for qubit regularization while preserving gauge invariance. This basis is the well-known \hbox{\ac{irrep}} basis of the $\SU(N)$ gauge group. A systematic approach to constructing this basis was outlined in \cite{Liu:2021tef}, which we use to develop a pictorial representation of the basis states as a tensor network of monomer and dimer tensors. This \ac{MDTN} basis not only provides a natural framework for qubit regularization but also enables the construction of new types of lattice gauge theories beyond the widely studied \ac{KS} Hamiltonians often explored in this context \cite{Kogut:1974ag}.

Qubit regularization of asymptotic freedom in gauge theories is likely to be trickier than in the $\dm=1$ spin model examples that we mentioned above, especially since gauge symmetries are not symmetries acting on the physical Hilbert space of the theory, but are rather constraints defining the physical Hilbert space. As we study this new class of theories, we may encounter different situations. One possibility is that the universality class of the fixed points of the \ac{RG} flow may depend on the chosen qubit regularization scheme. Another possibility is that the fixed point that governs the original \ac{QFT} may still be present in the \ac{RG} flow diagram, but can only be reached through a careful choice and tuning of the microscopic lattice model. While there is a lot of recent work that explores simple truncated lattice gauge theories on quantum computers \cite{Byrnes:2005qx,Ciavarella:2023mfc,Farrell:2024fit,Zemlevskiy:2024vxt,Crippa:2024cqr,Peng:2024xbl}, questions about quantum critical points in these qubit regularized theories and how the continuum limits can emerge continues to be an active area of research \cite{Berenstein:2022wlm,Hashizume:2021qbb}.
A systematic exploration, especially in two and three spatial dimensions, will be challenging even with powerful new techniques like the tensor-network methods. 

Since our motivation stems from the study of RG flows in qubit-regularized theories, we consider the popular \ac{KS} Hamiltonian \cite{Kogut:1974ag} as just one of several possible formulations to explore. The \ac{MDTN} basis of \acp{LGT} enables the construction of sign-problem-free quantum Hamiltonians, allowing these models to be studied using standard \ac{MCMC} methods. Here, we demonstrate that these new Hamiltonians successfully capture key qualitative features of traditional gauge theories, such as confinement, deconfinement, and finite-temperature transitions between these phases. Additionally, we find that a simple plaquette operator—analogous to that in the \ac{KS} Hamiltonian but constructed within the \ac{MDTN} basis—can lower the string tension, potentially inducing quantum phase transitions where asymptotic freedom may emerge.

Our work is organized as follows. In \cref{sec-2}, we construct the orthonormal basis of the physical Hilbert space for traditional $\SU(N)$ lattice gauge theories based on the \hbox{\acp{irrep}} of the gauge group. We show that the elements of this basis can be viewed as a network of monomer and dimer tensors, which we refer to as the \ac{MDTN} basis. Qubit regularization then becomes a straightforward step in this basis. In \cref{sec-3}, we construct a new class of qubit regularized gauge theory Hamiltonians using the \ac{MDTN} basis, consisting of two terms: a term that is diagonal in the \ac{MDTN} basis, and hence is essentially classical, and a noncommuting term that introduces quantum fluctuations in the theory. Leveraging the similarity between our qubit-regularized gauge theories and the transverse field Ising model, in \cref{sec-4}, we define classical gauge theories, similar to the classical Ising model, that we argue can capture the confinement-deconfinement physics of traditional gauge theories at finite temperatures. In \cref{sec-5}, using \ac{MCMC} methods, we confirm this expectation by reproducing the well-known confinement-deconfinement results at finite temperatures in $\SU(2)$ and $\SU(3)$ gauge theories. In particular, we verify the universality of these transitions in both $\dm=2$ and $\dm=3$. In \cref{sec-6}, we demonstrate how the term that induces quantum fluctuations in our Hamiltonian has the ability to lower the string tension of the classical term in the confined phase. We achieve this by studying the ground state of an $\SU(2)$ plaquette chain with two heavy matter particles. Finally, in \cref{sec-7}, we conclude by summarizing our findings and outlining a vision for future research.  

\section{Monomer-Dimer Tensor-Network}
\label{sec-2}

One of the key insights provided by qubit regularization in \acp{LGT} is the interpretation of the physical Hilbert space through the lens of \acp{irrep} of the gauge symmetry (see Ref.~\cite{Liu:2021tef}). While the \ac{irrep}-based approach was fundamental to the D-theory framework\cite{Chandrasekharan:1996ih, Brower:1997ha} and has been widely used in recent studies of gauge theories motivated by quantum computation~\cite{Raychowdhury:2019iki,Kadam:2022ipf,Burbano:2024uvn,Ciavarella:2024fzw}, we argue in this work that it offers a fundamentally new perspective on \acp{LGT}---one that extends beyond its original motivation in qubit regularization and quantum computation. In this section, we construct the \ac{irrep} basis and provide a pictorial representation in the form of a monomer-dimer tensor network (\ac{MDTN}). In the next section, we utilize this framework to construct new Hamiltonians for \acp{LGT}.

In order to understand the \ac{irrep} basis, let us begin with the full Hilbert space of traditional $\SU(N)$ \acp{LGT}, \( \cH^{\rm Trad} \), in the Hamiltonian formulation, before imposing Gauss's law. This full Hilbert space is the direct product of Hilbert spaces \( \cH^{\rm Trad}_\ell \) on the oriented links $\ell$ of the lattice and \( \cH^{\rm Trad}_s \) on the sites of the lattice:
\begin{align}
\cH^{\rm Trad} = \bigotimes_\ell \cH^{\rm Trad}_\ell \bigotimes_s \cH^{\rm Trad}_s\,.
\end{align}
Each link Hilbert space \( \cH^{\rm Trad}_\ell \) is associated with gauge degrees of freedom that describe a quantum particle moving on the surface of the \(\SU(N) \) manifold, while each site Hilbert space \( \cH^{\rm Trad}_s \) corresponds to matter degrees of freedom that transform according to some representation of the gauge group.

Traditionally, \( \cH^{\rm Trad}_\ell \) is constructed using the orthonormal ``position basis'' \( \ket{g} \), where \( g \) represents a point on the \( \SU(N) \) manifold. However, as explained in Ref.~\cite{Liu:2021tef}, for qubit regularization, it is often more useful to construct \( \cH^{\rm Trad}_\ell \) using a ``momentum basis", which in this case is the orthonormal \ac{irrep} basis of the gauge symmetry. 

If \( \lambda \) labels an \ac{irrep} of \(\SU(N) \), and \( V_\lambda \) denotes the corresponding Hilbert space with dimension \( d_\lambda \), then by the Peter-Weyl theorem, we can express~\cite{Liu:2021tef} 
\begin{align}
    \cH^{\rm Trad}_\ell \ =\  \bigoplus_{\lambda_\ell}\  V_{\lambda_\ell} \otimes V_{\bar{\lambda}_\ell}\,,
    \label{eq:ghs}
\end{align}
where \( V_{\lambda_\ell} \otimes V_{\bar{\lambda}_\ell} \) forms a \( d_{\lambda_\ell}^2 \)-dimensional subspace spanned by the orthonormal basis states \( | D^{\lambda_\ell}_{ij}\rangle \), with \( i, j = 1,2,\dots,d_{\lambda_\ell} \). Each \(\SU(N) \) \ac{irrep} \( \lambda_\ell \) appears exactly once in the direct sum. Here, \( \bar{\lambda}_\ell \) denotes the conjugate representation of \( \lambda_\ell \).

\begin{figure}[t]
\begin{center}
\includegraphics[width=0.35\textwidth]{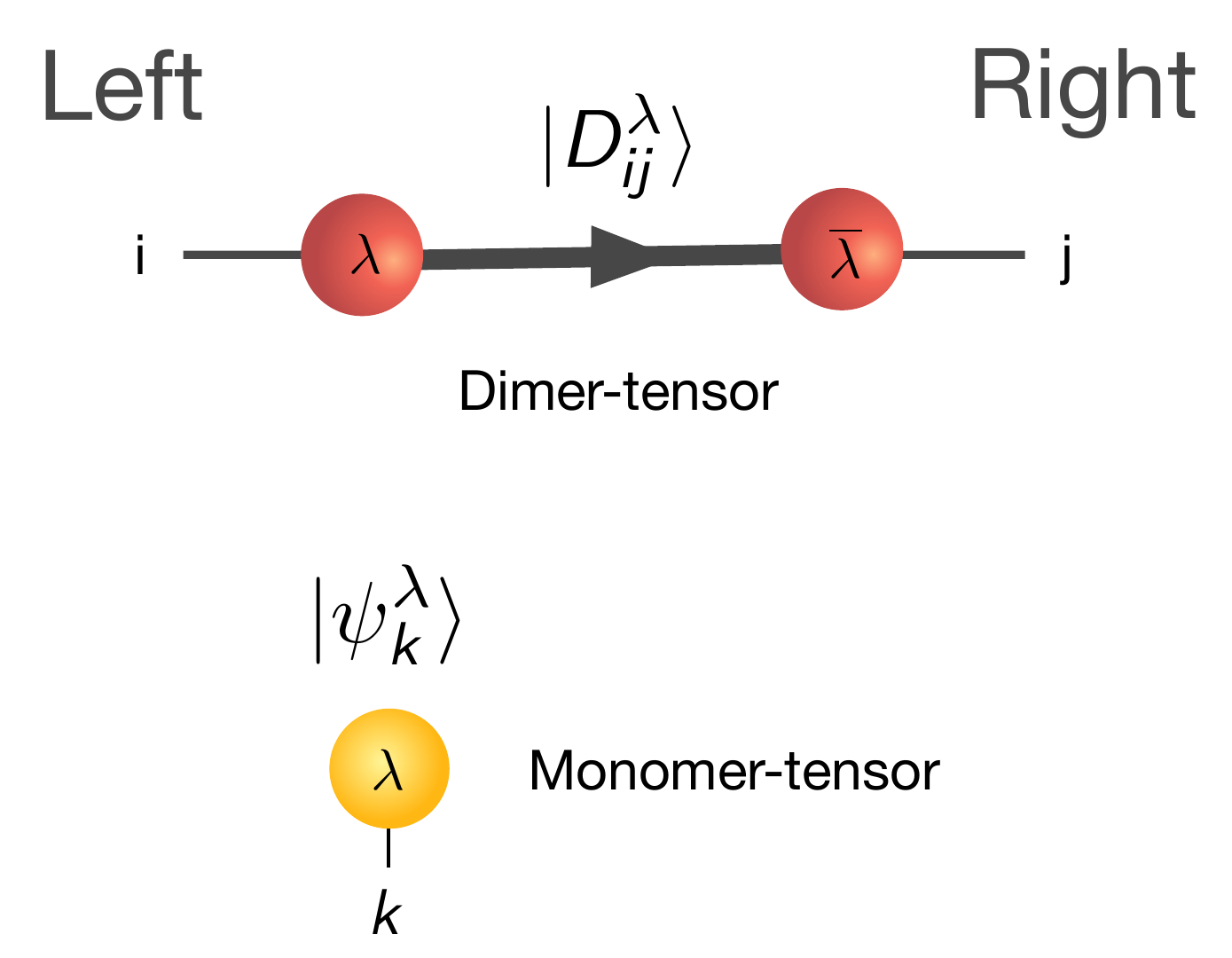}
\end{center} 
\caption{A pictorial representation of the two kinds of tensors that form the  \ac{MDTN} basis. The top shows the oriented dimer-tensor associated with the links, while the bottom shows the monomer-tensor at sites.}
\label{fig:mdt}
\end{figure}

In the orthonormal basis states \( | D^{\lambda_\ell}_{ij}\rangle \) associated with a link oriented from left to right, the index \( i \) corresponds to the degrees of freedom transforming under \( V_{\lambda_\ell} \) at the left lattice site, while \( j \) corresponds to those transforming under \( V_{\bar{\lambda}_\ell} \) at the right lattice site. The basis states \( | D^{\lambda_\ell}_{ij}\rangle \) at each link can be collectively interpreted as a tensor associated with an oriented dimer, labeled by \( \lambda_\ell \). A pictorial representation of this dimer-tensor is illustrated in \cref{fig:mdt} (top).

In contrast to the link Hilbert space, the site Hilbert space \( \cH^{\rm Trad}_s \) is associated with matter degrees of freedom that transform according to some representation of the gauge group and can naturally be written as  
\begin{align}
    \cH^{\rm Trad}_s \ =\  \bigoplus_{\lambda_s} \ V_{\lambda_s}\,,
    \label{eq:mhs}
\end{align}  
where each \ac{irrep} label \( \lambda_s \) corresponds to the representation \( V_{\lambda_s} \), which is spanned by the basis states \( | \psi^{\lambda_s}_k\rangle \) for \( k = 1,2,\dots,d_{\lambda_s} \). Collectively, these basis states can be interpreted as a tensor associated with a monomer on the site, labeled by \( \lambda_s \). A pictorial representation of this monomer-tensor is also illustrated in \cref{fig:mdt} (bottom).

For every fixed set of dimer-tensors and monomer-tensors, labeled as  
\( \{\lambda_\ell\} \) and \( \{\lambda_s\} \) respectively, one obtains the subspace  
\begin{align}
\cH^{\mdc} = \bigotimes_\ell  \left( V_{\lambda_\ell} \otimes V_{\bar{\lambda}_\ell} \right) \bigotimes_s V_{\lambda_s}
\end{align}  
of the full traditional Hilbert space \( \cH^{\rm Trad} \). However, the physical Hilbert space of the \ac{LGT}, \( \cH_\phys \), is obtained  
by projecting \( \cH^{\mdc} \) onto the space of gauge-invariant states, thereby imposing Gauss’s law.

Gauge transformations act on the oriented dimer-tensors (i.e., \( | D_{ij}^{\lambda_\ell}\rangle \) ) depending on whether the transformation is associated with the left site or the right site. In contrast, the monomer-tensors (i.e., \( | \psi^{\lambda_s}_k\rangle \) ) are associated with sites and transform as \( V_{\lambda_s} \). 

\begin{figure}[t]
\begin{center}
\includegraphics[width=0.35\textwidth]{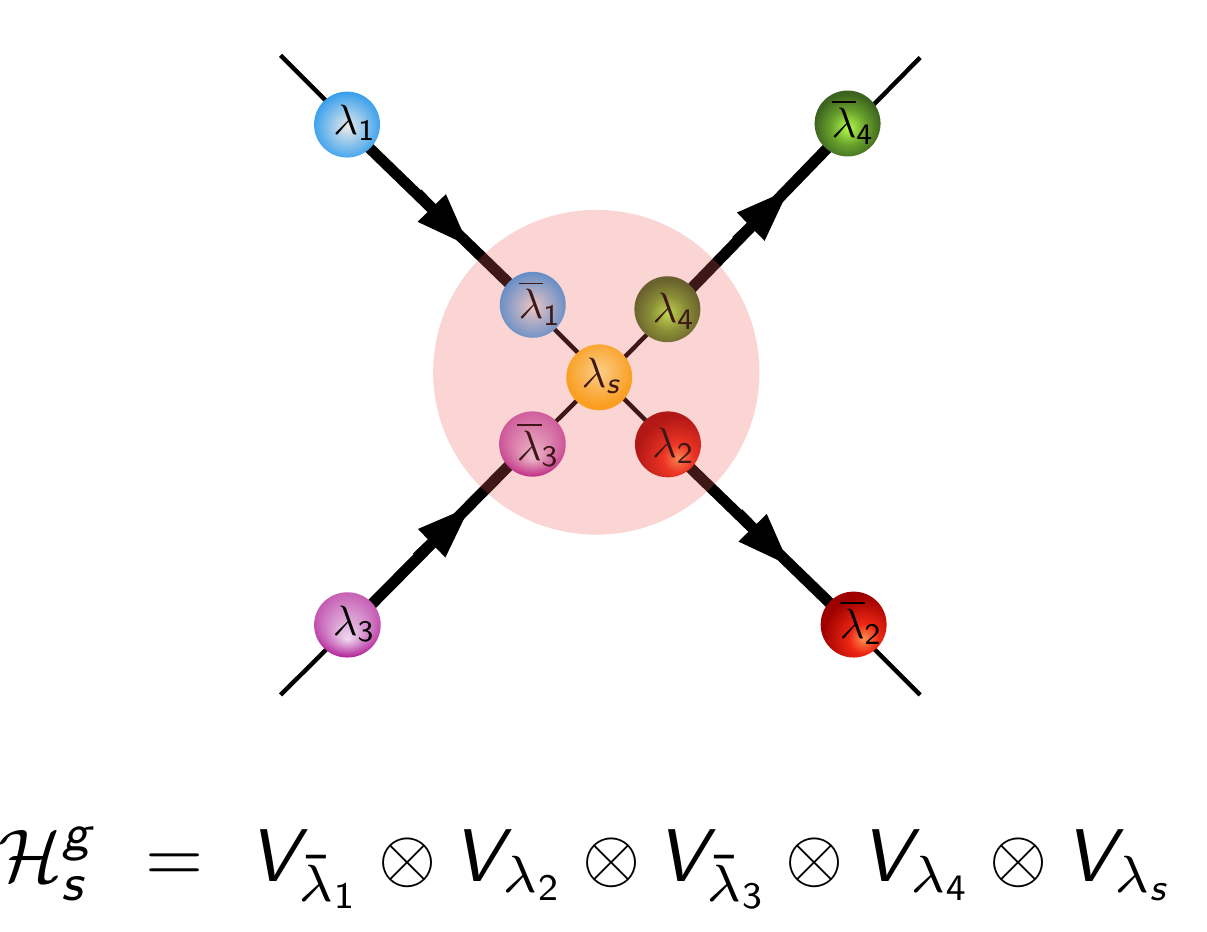}
\end{center} 
\caption{A pictorial representation of $\Hsg$ involving four dimer-tensors and a monomer-tensor. In this illustration, $\Hsg=V_{\bar{\lambda}_1}\otimes V_{\lambda_2}\otimes V_{\bar{\lambda}_3}\otimes V_{\lambda_4} \otimes V_{\lambda_s}$ and the physical Hilbert space is obtained by projecting on to $\SU(N)$ singlets of $\Hsg$. The various singlets are labeled by $\alpha_s = 1,2,\dots,\ldim$.}
\label{fig:hsp}
\end{figure}

From these transformation properties, we observe that gauge transformations at a lattice site \( s \) act on the Hilbert space \( \mathcal{H}_s^g \), which is the direct product of all \acp{irrep} \( V_{\lambda} \) associated with the dimer-tensors and monomer-tensors at that site. An illustration of \( \mathcal{H}_s^g \) is shown in \cref{fig:hsp}. Imposing Gauss's law involves constructing the singlet subspace of \( \mathcal{H}_s^g \), whose dimension we denote as \( \mathcal{D}(\mathcal{H}_s^g) \). An orthonormal basis for this singlet space is obtained through appropriate tensor contractions (or fusion rules) on the indices of the \( V_{\lambda} \)'s within \( \mathcal{H}_s^g \). Instead of specifying the precise form of these contractions, we simply label the orthonormal basis states at each site using an index \( \alpha_s = 1, 2, \dots, \mathcal{D}(\mathcal{H}_s^g) \).

The set of dimer-tensors and monomer-tensors, labeled by \( \{\lambda_\ell\} \) and \( \{\lambda_s\} \) respectively, along with the indices \( \{\alpha_s\} \) that label the gauge-invariant states at lattice sites, forms an orthonormal basis for a traditional \( \SU(N) \) \ac{LGT}. These states can be pictorially represented as \ac{MDTN}, which we denote as \( \ket{\mdck} \).

While the dimension of the physical Hilbert space in \acp{LGT} with discrete gauge groups can be computed exactly \cite{Mariani:2023eix}, the calculation becomes more challenging for continuous gauge groups. However, the \ac{MDTN} basis states provide a useful framework for making progress. For a given set of monomer-dimer tensors, which we denote as \( \mdc \) without specifying the labels \( \alpha_s \), the dimension of \( \cH_\phys \) can be expressed in closed form as  
\begin{align}
\dim(\cH_\phys) = \prod_s \ldim\,.
\label{eq:cHpdim}
\end{align}
In a traditional gauge theory, the total dimension of the physical Hilbert space is given by the formal expression  
\begin{align}
\dim(\cH^{\rm Trad}_\phys) = \sum_{\mdc} \prod_s \ldim\,,
\label{eq:cHTpdim}
\end{align}  
which, due to the unrestricted sum over the representations \hbox{\(\{\lambda_s\}\) and \(\{\lambda_\ell\}\)}, is infinite even on a finite lattice. The goal of qubit regularization is to make this dimension finite.

Qubit-regularized \acp{LGT} can be naturally constructed using the \ac{MDTN} basis states by restricting the allowed values of \( \lambda_\ell \) and \hbox{\(\lambda_s\)} in \hbox{\(\ket{\mdck}\)}. With this restriction, the dimension of the physical qubit-regularized Hilbert space, \( \cH^Q_\phys \), is given by  
\begin{align}
\dim(\cH^Q_\phys) = \sum^\prime_{\mdc} \prod_s \ldim\,,
\label{eq:cHQpdim}
\end{align}  
which is finite because the sum runs over a restricted set of monomer-dimer configurations, as indicated by the prime symbol.  

A simple qubit regularization for \(\SU(N) \) gauge theories was introduced in \cite{Liu:2021tef}, which restricts \hbox{\( \lambda_{\ell}\)} and \hbox{\(\lambda_s \)} to anti-symmetric \acp{irrep} corresponding to single-column Young diagrams. We refer to this as the \ac{ASQR}. In this scheme, the allowed representations for both \hbox{\(\lambda_\ell\)} and \hbox{\(\lambda_s\)} are \hbox{\(\{1,2\}\)} for \(\SU(2) \) and \hbox{\(\{1,3,\bar{3}\}\)} for \(\SU(3) \) \ac{LGT}.  

\begin{figure}[t]
\centering
\includegraphics[width=0.48\textwidth]{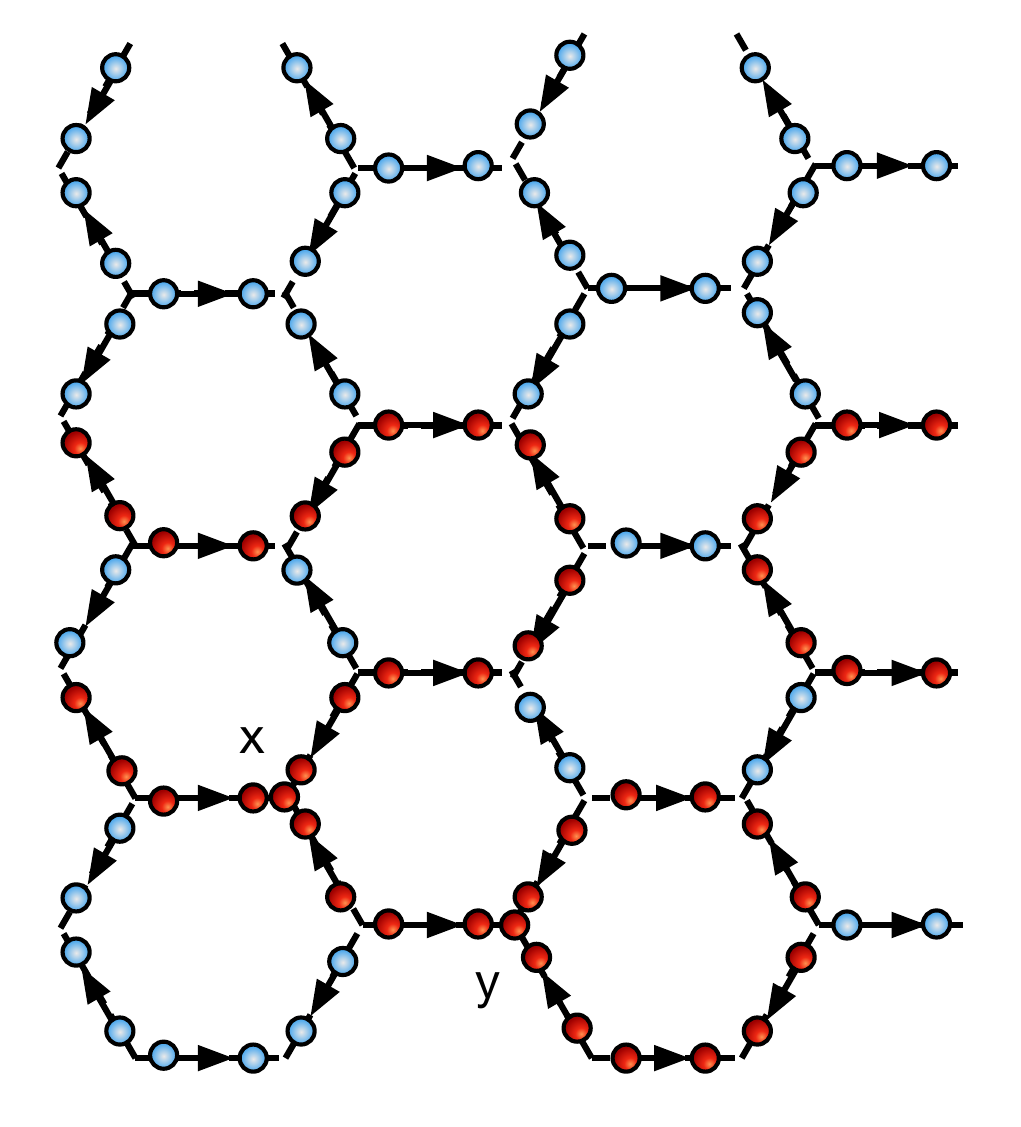}
\caption{Illustration of a \hbox{\ac{MDTN}} configuration in a qubit-regularized $\SU(2)$ pure gauge theory within the \ac{ASQR} scheme, with nondynamical source matter fields located at sites $x$ and $y$. Blue circles represent $\lambda=1$ (singlets), while red circles denote $\lambda=2$ (doublets). Monomer tensors in the trivial \ac{irrep} are not shown, whereas those in the doublet \ac{irrep} are displayed at sites $x$ and $y$. On all sites, ${\cal D}_s(\cH_s^g) = 1$, except at sites $x$ and $y$, where ${\cal D}_s(\cH_s^g) = 2$. The label $\alpha_s$ is suppressed in this illustration.}
\label{fig:su2Qc}
\end{figure}

\begin{figure}[t]
\centering
\includegraphics[width=0.48\textwidth]{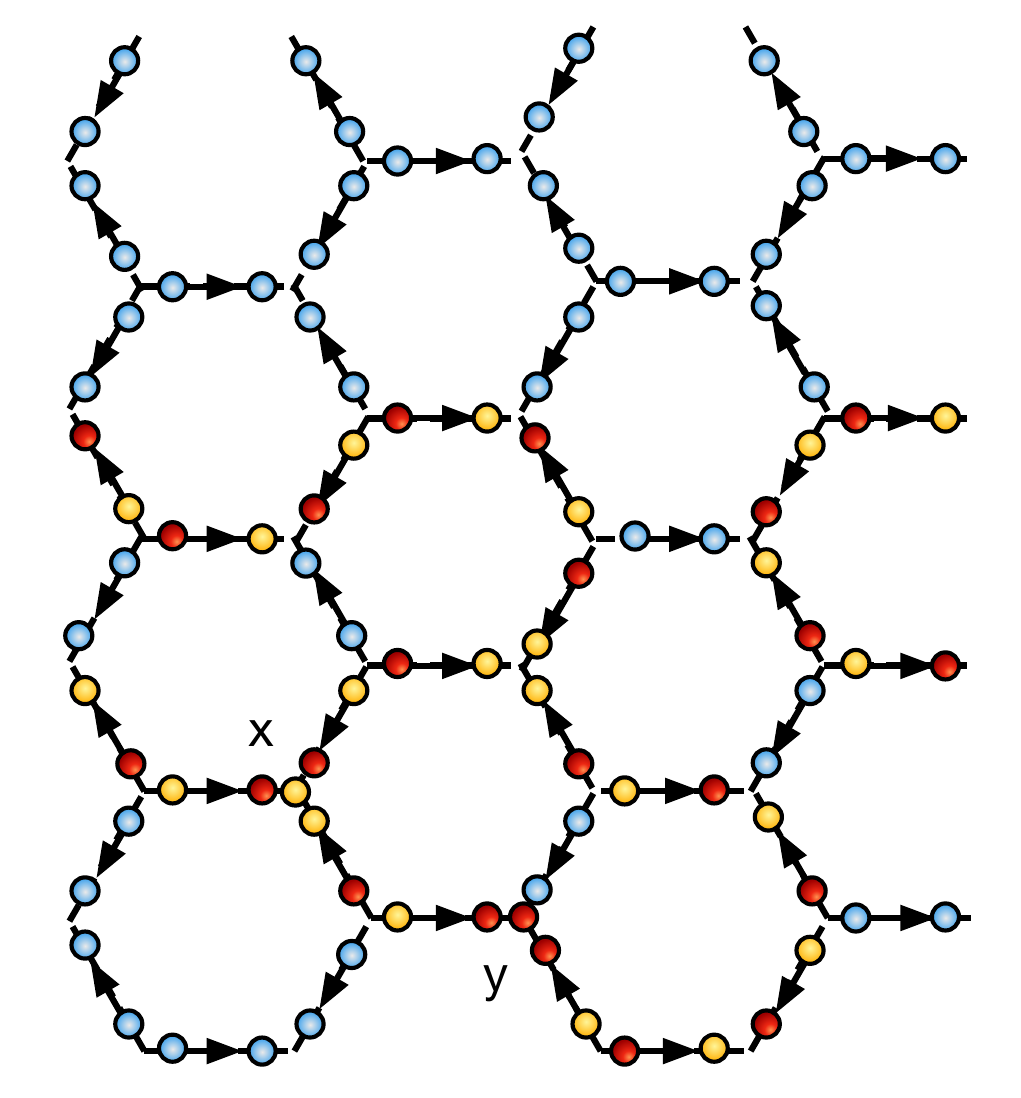}
\caption{Illustration of a \hbox{\ac{MDTN}} configuration in a qubit-regularized $\SU(3)$ pure gauge theory within the \ac{ASQR} scheme, with nondynamical source matter fields located at sites $x$ and $y$. Blue circles represent $\lambda=1$ (singlets), while red and yellow circles denote $\lambda=3$ (triplets) and $\lambda=\bar{3}$ (anti-triplets), respectively. Monomer tensors in the trivial \ac{irrep} are not shown, whereas those in the triplet and anti-triplet \acp{irrep} are displayed at sites $x$ and $y$. On all sites, ${\cal D}_s(\cH_s^g) = 1$, except at site $x$, where ${\cal D}_s(\cH_s^g) = 2$. The label $\alpha_s$ is suppressed in this illustration.
}
\label{fig:su3Qc}
\end{figure}

\Cref{fig:su2Qc} illustrates an \(\SU(2) \) monomer-dimer configuration on a honeycomb lattice, where all sites have \( \lambda_s=1 \) except for two sites, \( x \) and \( y \), which host nondynamical source matter fields with \( \lambda_s=2 \). Similarly, \cref{fig:su3Qc} shows an \(\SU(3) \) configuration, where site \( x \) has a matter field in the \( \lambda_s=\bar{3} \) \ac{irrep}, while site \( y \) hosts a matter field in the \( \lambda_s=3 \) \ac{irrep}.   

In a primitive form, \ac{MDTN} basis states have appeared in the condensed matter literature as monomer-dimer models, where they have been linked to \( \mathbb{Z}_2 \) and \( \mathrm{U}(1) \) gauge theories \cite{Moessner:2000ahy,Chen:2009vza,moes2008}.

\section{Qubit Regularized Gauge Theories}
\label{sec-3}

Since the \ac{MDTN} basis states \( \ket{\mdck} \) form an orthonormal basis of the physical Hilbert space \( \cH^{\rm Q}_\phys \) of qubit-regularized \acp{LGT}, a variety of local Hamiltonians can immediately be constructed by defining matrix elements in this basis. The traditional \ac{KS} Hamiltonian \cite{Kogut:1974ag} is just one such possibility. However, a key challenge with analyzing the \ac{KS} Hamiltonian is that the matrix elements of its plaquette terms depend on Clebsch-Gordan coefficients and other complex group-theoretic factors \cite{Zohar:2021nyc,Halimeh:2024bth}. This often leads to sign problems, making it difficult to apply \ac{MCMC} methods to study qubit regularization effects, particularly in higher dimensions. While sign problems may not pose an issue for quantum computation, gaining deeper insight into qubit-regularized gauge theories requires the ability to study their physics on large systems not accessible to quantum computations today. This motivates the exploration of new Hamiltonians, beyond those proposed by \ac{KS}, inspired by the \ac{MDTN} basis states.

The \ac{KS} Hamiltonian is often regarded as the key Hamiltonian to study in \ac{LGT} \cite{Mezzacapo:2015bra,Muller:2023nnk}, though much of its motivation is the fixed point amenable to perturbation theory. In the traditional formulation, the infinite-dimensional Hilbert space allows for an exact solution of the \ac{KS} Hamiltonian in the limit of vanishing gauge coupling. In this limit, the Hamiltonian describes free gluons, and the lattice theory flows to the Gaussian \ac{UV} fixed point of the continuum Yang-Mills theory. However, with qubit regularization, this exact solvability—particularly at weak couplings—is no longer available. In fact, it would indeed be surprising if the infinite-dimensional local Hilbert space of the continuum theory could be recovered simply by taking the perturbative weak-coupling limit of a qubit-regularized theory with a strictly finite-dimensional local Hilbert space. 

The above discussion suggests that the \ac{KS} Hamiltonian may not be as central as traditionally assumed, making the exploration of alternative Hamiltonians worthwhile. Fortunately, using the \ac{MDTN} basis states, we can systematically construct a broad class of \acp{LGT} that serve as natural starting points for studying qubit-regularized gauge theories. The focus of these explorations is to determine whether continuum gauge theories can emerge at critical points beyond the reach of perturbation theory.

\begin{figure*}[t]
\centering
\includegraphics[width=0.95\textwidth]{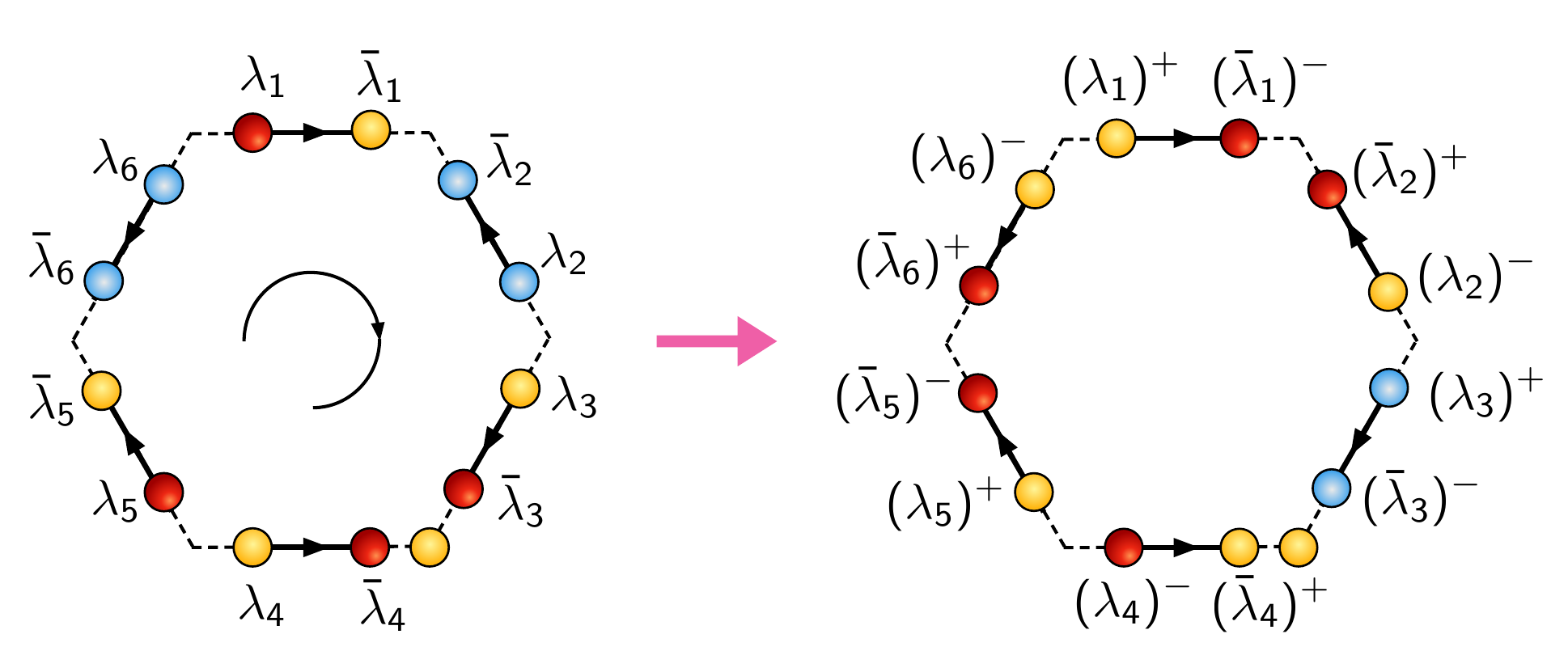}
\caption{Illustration of the action of $\chUp$. Here, we assume that the states $\ket{\mdck}$ are restricted according to the \ac{ASQR} scheme, where $\lambda^+$ denotes the irrep obtained by adding a box to the Young tableau of $\lambda$, while $\lambda^-$ represents the \ac{irrep} obtained by removing a box, following the cyclic property of modulo \(N\)-boxes. The orientation of the plaquette, shown on the left, sets the convention for how the \acp{irrep} change. Notably, this construction preserves the \(N\)-ality. Thus, every plaquette action generates allowed states in the physical Hilbert space, although the singlet spaces and their dimensions before and after the action may differ. The complete definition of how singlet spaces are mapped is encoded in the coefficients \(c(\mdck,\{\alpha_s'\},P)\), introduced in the definition of $\chUp$ in \cref{eq:chupdef}.  
\label{fig:plaqop}}
\end{figure*}

In this work, we propose a simple \ac{LGT} Hamiltonian of the form  
\begin{align}
H_Q \ =\ \sum_\ell \chE_\ell - \delta \sum_{P} \Big(\chUp + \chUp^\dagger\Big),
\label{eq:cHQmod}
\end{align}
where the first term is a sum of local link operators, $\chE_\ell$, such that the \ac{MDTN} basis states $\ket{\mdck}$ form an eigenbasis of these operators with eigenvalues given by ${\cal E}(\lml)$, i.e.,
\begin{align}
\chE_\ell\ket{\mdck} = {\cal E}(\lml)\, \ket{\mdck}.
\end{align}
The second term is a sum over operators, $(\chUp +\chUp^\dagger)$, defined on lattice plaquettes, similar in spirit to the \ac{KS} Hamiltonian. When $\chUp$ acts on $\ket{\mdck}$, it changes the $\SU(N)$ irreps $\{\lambda_\ell\}$ on the links of the plaquette \(P\) to $\{\lambda_\ell'\}$. It also modifies the singlet states on the vertices of the plaquette, transforming \(\alpha_s \to \alpha'_s\) according to specific rules. These rules are encapsulated in the following definition:  
\begin{align}
\chUp &= \sum_{\substack{\dck\\ \dckp},P} c\Big(\dck,\dckp,P\Big) \times{}\nonumber\\
&\qquad\qquad\ket{\{\lambda_s\},\dckp}\bra{\{\lambda_s\},\dck} 
\label{eq:chupdef}
\\ 
\chUp^\dagger &= \sum_{\substack{\dck\\ \dckp},P} c^*\Big(\dck,\dckp,P\Big) \times{}\nonumber\\
&\qquad\qquad\ket{\{\lambda_s\},\dck}\bra{\{\lambda_s\},\dckp}
\label{eq:chupdagdef}
\end{align}
where the complex coefficients $c$ give the matrix elements between gauge-invariant states related by changes occurring on the plaquette $P$ due to the action of $\chUp$. We allow the coefficients $c$ to depend on the location \hbox{\(P\)} of the plaquettes, but to maintain translational and rotational invariance, this dependence needs to be appropriately periodic.

We emphasize that while $\chUp$ is very similar to the plaquette operators in the \ac{KS} Hamiltonian, its explicit representation in terms of the traditional link operators \(\chU\) used in \acp{LGT} is nontrivial.\footnote{These traditional link operators \(\chU\) form a complete basis for the local operator algebra~\cite{Liu:2021tef}, and hence $\chUp$ can be expressed in terms of them.} For this reason, it is preferable not to rewrite \(\chUp\) in terms of \(\chU\), and work directly with the coefficients \hbox{\(c\)} given in \cref{eq:chupdef}. In the large $N$ limit, similar simplifications of the plaquette operators seem to emerge naturally \cite{Ciavarella:2024fzw}.

While \cref{eq:cHQmod} defines a valid Hamiltonian even in the traditional infinite-dimensional Hilbert space, it is particularly useful in the context of qubit regularization. In this work, we employ it within the \ac{ASQR} scheme. 

Importantly, if \(\delta \geq 0\) and the coefficients $c$ are real and non-negative, all off-diagonal terms in \cref{eq:cHQmod} are non-positive. Consequently, the quantum \ac{LGT} remains free of sign problems  ~\hbox{\cite{Marshall55}}, allowing its physics to be studied using well-known Hamiltonian \ac{MCMC} methods.

\section{Classical Lattice Gauge Theories}
\label{sec-4}

Can qubit-regularized \acp{LGT}, such as the one introduced in the previous section in \cref{eq:cHQmod}, recover the physics of continuum gauge theories, including asymptotic freedom? This is a challenging question to answer, as perturbative methods commonly used to analyze traditional lattice gauge theories are no longer applicable near the critical points of qubit-regularized \acp{LGT}. 
Moreover, while Hamiltonian Monte Carlo methods may be employed to study the physics described by \cref{eq:cHQmod} for certain classes of plaquette operators, as discussed in the previous section, identifying quantum critical points and analyzing their universality classes will require significant effort.
A much simpler question that we can address more easily is whether the Hamiltonian in \cref{eq:cHQmod} exhibits at least the qualitative physics of confinement and deconfinement. For example, can it reproduce some of the features of traditional lattice gauge theories related to finite-temperature confinement-deconfinement transitions? 

Finite-temperature physics is often more accessible because the quantum nature of physical systems becomes less significant at temperatures much higher than the energy level spacing of a theory. As a result, finite-temperature critical phenomena can often be captured by classical statistical mechanics. 
As an example, consider the finite-temperature Ising phase transition starting from the transverse field quantum Ising model:
\begin{align}
H_{\rm TFI} \ =\ - \sum_{\langle ij\rangle} S^z_i S^z_j + \delta \sum_i S^x_i\,,
\label{eq:Ising}
\end{align}
where the first term alone results in a classical Ising model, while the second term introduces quantum fluctuations through a transverse field. However, to study the finite-temperature Ising transition, we can set $\delta=0$ and analyze just the classical model. The quantum fluctuations introduced by the $\delta$ term are irrelevant for capturing the universal finite-temperature critical behavior.

Note that the Ising Hamiltonian \cref{eq:Ising} bears a striking resemblance to the gauge theory Hamiltonian \cref{eq:cHQmod}. The first term of the gauge theory Hamiltonian can also be interpreted as a classical Hamiltonian in the \ac{MDTN} basis, while the second term introduces quantum fluctuations. Drawing on this analogy with the Ising model, we propose a class of ``classical lattice gauge theories'' defined by Hamiltonians whose eigenbasis coincides with the \ac{MDTN} basis. One such Hamiltonian is given by  
\begin{align}
H^{\rm cl}_Q \ =\ \sum_\ell \chE_\ell\,,
\label{eq:cHcl}
\end{align}
obtained by setting $\delta=0$ in \cref{eq:cHQmod}. A natural question is whether \cref{eq:cHcl} can capture the universal features of finite-temperature phase transitions in traditional lattice gauge theories. This question can be addressed in any dimension using powerful loop algorithms~\cite{Adams:2003cca,Chandrasekharan:2006tz}, as we discuss below.

\begin{figure*}[t]
\centering
\includegraphics[width=0.49\textwidth]{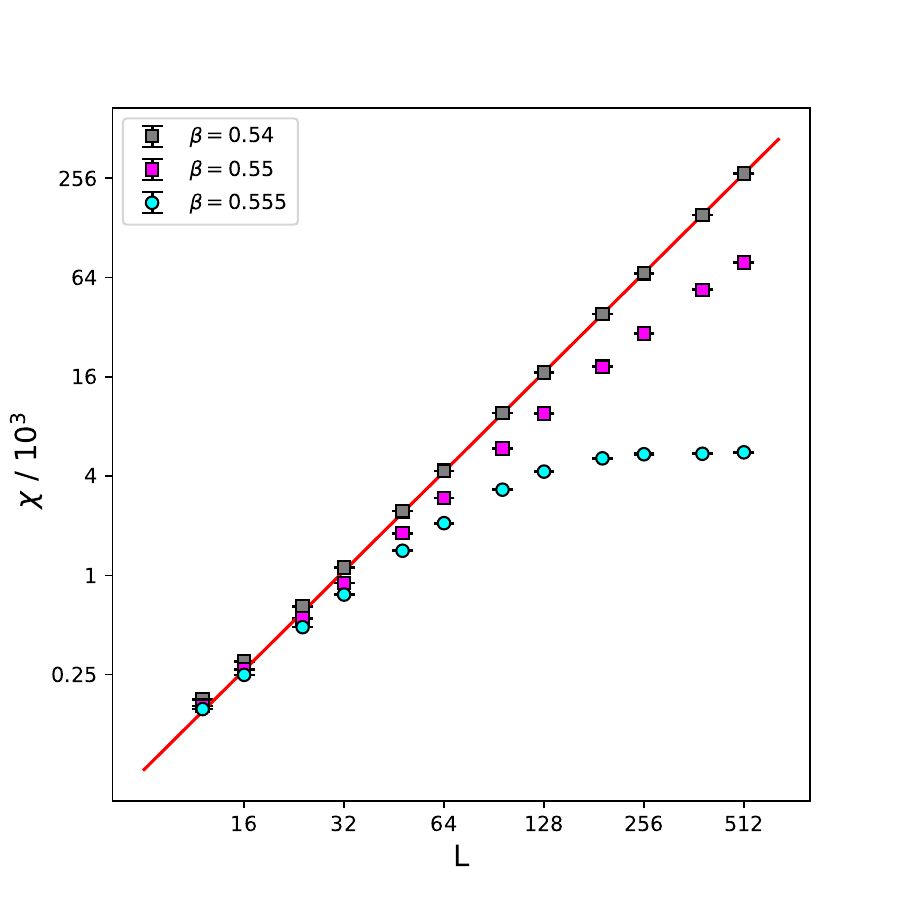}
\includegraphics[width=0.49\textwidth]{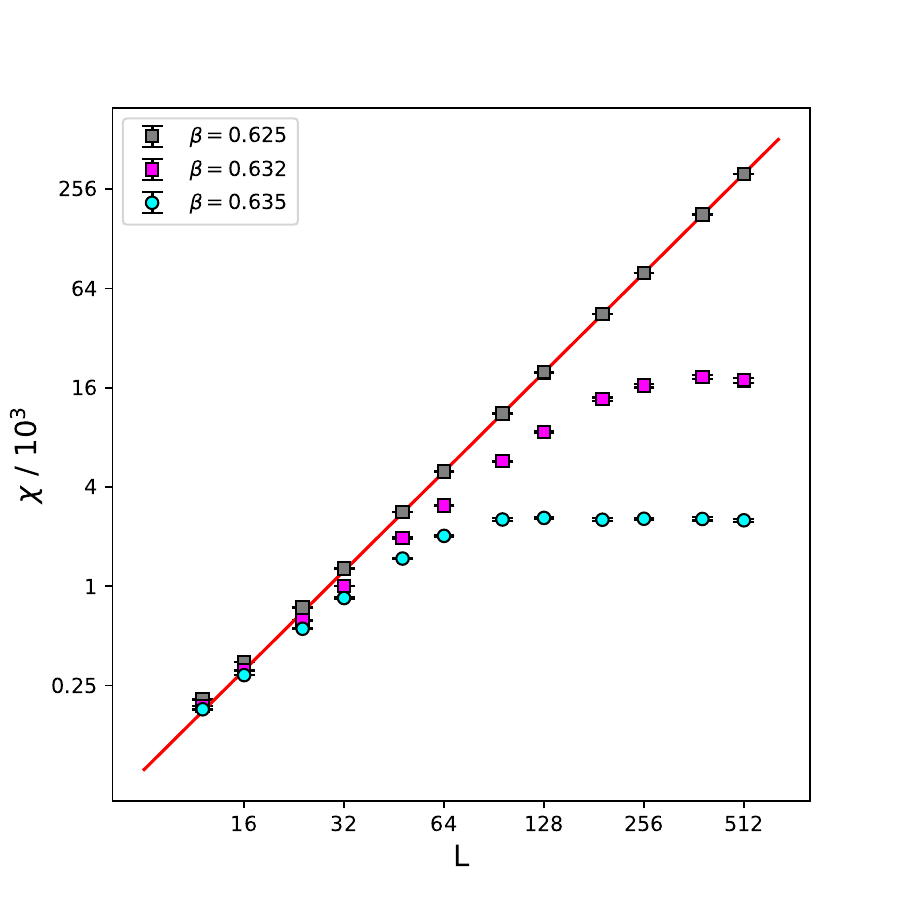}
\includegraphics[width=0.49\textwidth]{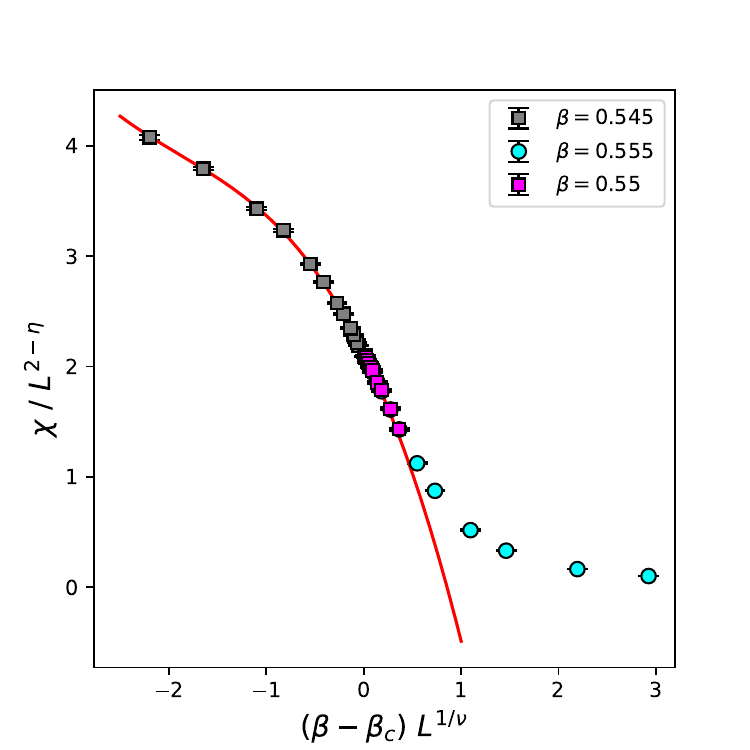}
\includegraphics[width=0.49\textwidth]{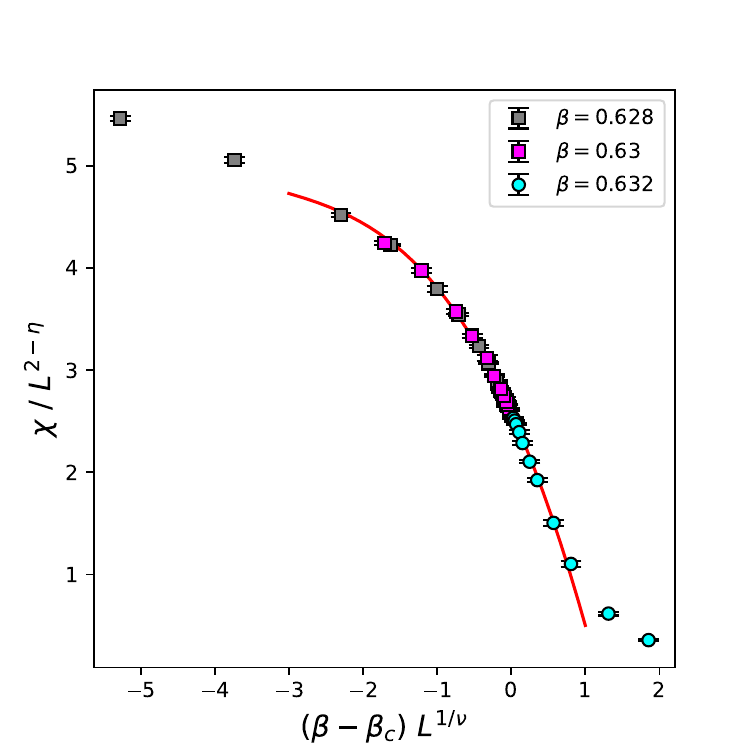}
\caption{Our \hbox{\ac{MCMC}} results for the susceptibility $\chi$ in $\dm=2$. In the top row we show the plots of $\chi$ as a function of $L$ at three different values of $\beta$ for $\SU(2)$ (left) and $\SU(3)$ (right). The straight lines are fits to the form $\chi \sim A L^2$. The fit is excellent if we use data with $L > 100$ and gives $A=1.039(2)$ and $A=1.210(2)$ for $\SU(2)$ and $\SU(3)$ respectively. These transitions are consistent with second order. In the bottom row we show our \hbox{\ac{MCMC}} data for $\chi$ scaled according to \cref{eq:scal} for $\SU(2)$ (left) and $\SU(3)$ (right). For $\SU(2)$ we use the two dimensional Ising critical exponents $\eta=1/4$ and $\nu=1$ with $\beta_c=0.54929$, while for $\SU(3)$ we use the 3-state Potts model critical exponents $\eta=4/15$ and $\nu=5/6$ with $\beta_c=0.63096$. The critical values of $\beta$ are obtained using a fit to \cref{eq:scal} assuming $f(x)$ is a fourth-order polynomial. In this fit, we only use data close to $x\approx 0$ and obtain the coefficients shown in the first two rows of \cref{tab:2dcf}. The fit functions are shown as lines in the two graphs in the bottom row. \label{fig:D2chi}}
\end{figure*}

\begin{figure*}[htb]
\centering
\includegraphics[width=0.49\textwidth]{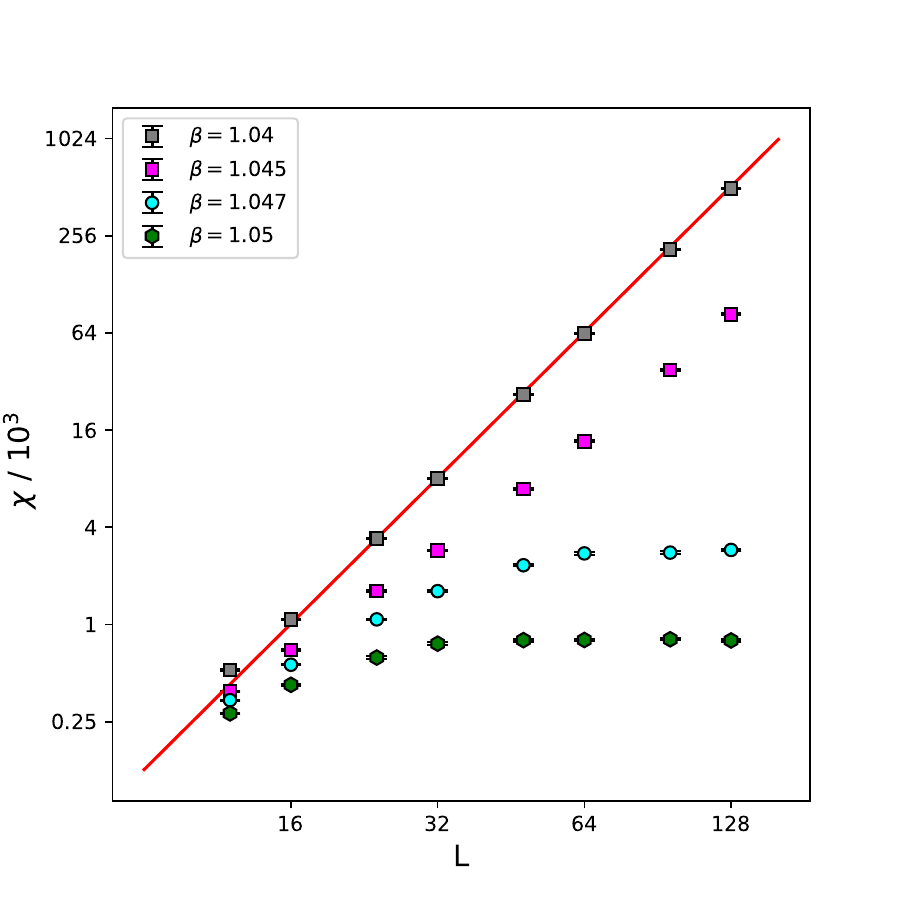}
\includegraphics[width=0.49\textwidth]{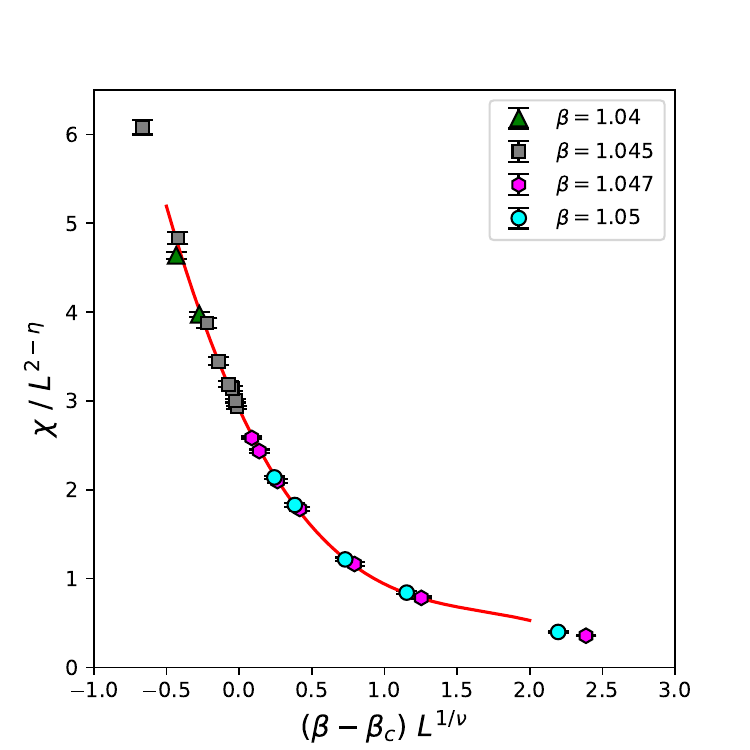}
\caption{Our \hbox{\ac{MCMC}} results for $\chi$ in $\dm=3$ for the $\SU(2)$ case. In the left plot, we show $\chi$ as a function of $L$ at four different values of $\beta$. The deconfined phase is clearly visible at $\beta=1.04$, where the data fits the form $\chi \sim A L^3$ when using data with $L > 32$, yielding $A=0.240(1)$. The confined phases are also evident at both $\beta=1.047$ and $\beta=1.05$. The transition is consistent with being second order. In the right figure, we show our \hbox{\ac{MCMC}} data for $\chi$ scaled according to \cref{eq:scal}, using the three-dimensional Ising critical exponents $\eta=0.0366$ and $\nu=0.6298$, with $\beta_c=1.04530(1)$. The critical value of $\beta$ was obtained by fitting to \cref{eq:scal}, assuming $f(x)$ is a fourth-order polynomial. In this fit, we use only data near $x\approx 0$, obtaining the coefficients shown in the third row of \cref{tab:2dcf}. The fit function is represented by the line in the right graph.}
\label{fig:D3su2}
\end{figure*}

\section{Confinement-Deconfinement Transition}
\label{sec-5}

The finite-temperature confinement-deconfinement transition in $\dm=2$ and $\dm=3$ spatial dimensions has been extensively studied over the years using traditional lattice gauge theories. However, similar studies in qubit-regularized gauge theories, particularly in higher dimensions, remain scarce. Recently, a few studies in both Abelian and non-Abelian gauge theories have begun to emerge \cite{Banerjee:2013dda,Banerjee:2017tjn,PhysRevLett.130.071901}.

Here, we will focus on pure gauge theories, where we expect to observe a confined phase at low temperatures and a deconfined phase at high temperatures. Furthermore, it is expected that the universality class of the associated confinement-deconfinement transition can be understood through spin models that are invariant under center symmetry \cite{Svetitsky:1982gs,Lucini:2003zr}. In pure $SU(N)$ gauge theories, these are expected to be $\mathbb{Z}_n$ spin models, where the confined phase corresponds to the symmetric phase and the deconfined phase corresponds to the broken phase. Focusing on $N=2$ and $N=3$, the transitions in $\mathbb{Z}_2$ spin models can be second order in both $\dm=2$ and $\dm=3$ dimensions, while $\mathbb{Z}_3$ spin models exhibit second-order transitions only in $\dm=2$ dimensions and always show first-order transitions in $\dm=3$ dimensions \cite{Pelissetto:2000ek}. Extensive studies in traditional \acp{LGT} have confirmed these general expectations \cite{McLerran:1981pb,Yaffe:1982qf,Gavai:1982er,Kogut:1984sa,Curci:1984yw,Gross:1984pq,Das:1985up,Iwasaki:1991pc,Teper:1993gp,Datta:2009jn,Caselle:2011fy,Caselle:2011mn,Gavai:2023cxc}.

As proposed in the previous section, we wish to explore whether these universal features of finite-temperature confinement-deconfinement transitions can be reproduced using qubit-regularized classical \acp{LGT} given by \cref{eq:cHcl}. We focus on pure gauge theory by setting $\lambda_s = 1$ everywhere. If we label the corresponding \ac{MDTN} states as \hbox{\(\ket{\dck}\)}, the classical partition function is given by
\begin{align}
Z_Q \ &=\ \sum_{\dck}\ \prod_\ell e^{-\beta {\cal E}(\lambda_\ell)}
\nonumber \\
&= \sum_{\dc}\ \prod_\ell e^{-\beta {\cal E}(\lambda_\ell)}\ \prod_s \ \ldim\,,
\label{eq:cpf}
\end{align}
where in the last step we summed over $\{\alpha_s\}$ for a fixed set of dimer-tensors $\dc$. 

For simplicity, we will study the model with the link eigenvalues given by
\begin{align}
{\cal E}(\lambda_\ell) \ &=\ 1 - \delta_{\lambda_\ell,0}\,.
\end{align}
This essentially assigns a unit of energy to all non-trivial dimer-tensors, while the energy of trivial dimer-tensors vanishes. This ensures that the theory is in the confined phase at zero temperature, since test matter sources with fundamental (or antifundamental) \hbox{\(\lambda_s\)} will need to be accompanied by a string of $\lambda_\ell \neq 1$ dimer-tensors, which incurs an energy cost.

In order to quantitatively study the finite temperature confinement-deconfinement transition, we study the partition for the pure gauge theory, but with nondynamical matter fields at precisely two locations $x$ and $y$ in the \acp{irrep} $\lambda_x$ and $\lambda_y$ respectively. If we label these \ac{MDTN} states as $\dcxyk$, this partition function is given by
\begin{align}
Z_Q^{(x,y)}\ & =\ \sum_{[\dcxy]}\ \prod_\ell e^{-\beta {\cal E}(\lambda_\ell)} \prod_s \ldim\,,
\label{eq:cpfxy}
\end{align}
where we have already performed the sum over $\{\alpha_s\}$ like in \cref{eq:cpf}. We expect the ratio of this partition function to the one without the matter fields to behave as
\begin{align}
Z^{(x,y)}_Q/Z_Q \sim e^{-\beta F(x,y)}\,,
\end{align}
where $F(x,y)$ is the excess free energy of introducing two heavy particles located at $x$ and $y$ in the pure gauge theory. 

For large separations, in the confined phase we expect $F(x,y)$ to grow linearly for large \(|x-y|\), while in the deconfined phase it does not grow indefinitely. We can detect this difference in the behavior of $F(x,y)$ through the susceptibility in a finite lattice with volume $L^\dm$ defined as
\begin{align}
\chi\ =\ \frac{1}{L^\dm} \sum_{x,y} \frac{Z^{(x,y)}_Q}{Z_Q}\,.
\label{eq:chi}
\end{align}
Using the above expectations for $F(x,y)$ we can show that
\begin{align}
\lim_{L\rightarrow \infty}
\chi \ =\ \left\{ 
\begin{matrix}
\chi_0 & \qquad \mbox{confined phase} \\ \\
A \ L^\dm  & \qquad \mbox{deconfined phase}
\end{matrix} 
\right.
\label{eq:cdtrans}
\end{align}
Further, if the confinement-deconfinement phase transition is second order, we expect
\begin{align}
\lim_{L\rightarrow \infty}\ \chi \sim f_0 L^{2-\eta}    
\end{align}
where $\eta$ is the universal critical exponent associated with the transition. Since the partitions functions \cref{eq:cpf,eq:cpfxy} do not suffer from sign problems, we can use well-known loop cluster methods to generate the required dimer configurations \([\dc]\) and \([\dcxy]\) distributed according to their classical Boltzmann weights  \cite{Adams:2003cca,Chandrasekharan:2006tz}. These loop algorithms also allow us to measure $\chi$ easily. 

Using these loop algorithms we have computed $\chi$ in both $\dm=2$ and $\dm=3$ for several values of $\beta$ and $L$. In $\dm=2$ we use a honeycomb lattice while in $\dm=3$ we use the diamond lattice. The details these lattice geometries are explained in \cref{app-1}.

Let us first focus on our results in $\dm=2$. We find clear evidence for the expected confinement-deconfinement transitions as a function of $\beta$.  In \cref{fig:D2chi} we plot our results for $\chi$ as a function of $L$ for three different values of $\beta$ close to the transition for $\SU(2)$ (left plot) and $\SU(3)$ (right plot). Consistent with the expectations from \cref{eq:cdtrans}, we see a deconfined phase at $\beta=0.54$ and a confined phase at $\beta=0.555$ for $\SU(2)$. In the case of $\SU(3)$ we find $\beta=0.625$ is deconfined while $\beta=0.635$ is confined. 

In $\dm=2$, these transitions seem second order with the susceptibility satisfying the scaling form ~\hbox{\cite{Pelissetto:2000ek}}
\begin{align}
\chi = L^{2-\eta} f((\beta-\beta_c) L^{1/\nu})
\label{eq:scal}
\end{align}
near the critical point. In the case of $\SU(2)$ we expect the second order transition to satisfy the two-dimensional Ising universality class with $\eta=1/4$ and $\nu=1$~\hbox{\cite{Wu:1982ra}}. For $\SU(3)$, a second order transition is possible within the three-state Potts model universality class where $\eta=4/15$ and $\nu=5/6$~\cite{Wu:1982ra}. Near the transition, we find that our values of $\chi$ are indeed consistent with this scaling behavior if we assume $f(x)\ \approx\ f_0 + f_1 x + f_2 x^2 + f_3 x^3$. We fit our data to this form to extract $\beta_c$. We show our results in the bottom row of \cref{fig:D2chi}. The values of the four coefficients are shown in \cref{tab:2dcf}.

\begin{table}[H]
\centering
\renewcommand{\arraystretch}{1.4}
\setlength{\tabcolsep}{4pt}
\begin{tabular}{c|c|c|c|c}
\TopRule
 $\beta_c$ & $f_0$ & $f_1$ & $f_2$ & $f_3$  \\
\MidRule
0.54929(1) & 2.120(2) & -1.74(1) & -0.46(3) & 0.04(3) \\
0.63096(2) & 2.584(4) & -1.69(2)& -0.50(5) & -0.05(3)\\
1.04530(1) & 2.92(1) & -3.94(6) & 1.95(7) & -0.38(2) \\
\BotRule 
\end{tabular}
\caption{ The first four Taylor coefficients for the scaling function $f(x)$ and $\beta_c$ in \cref{eq:scal} for $\dm=2,\ \SU(2)$ (first row), $\dm=2,\ \SU(3)$ (second row) and $\dm=3,\ \SU(2)$ (third row). The $\chi^2/DOF \sim 1$ for all these fits. \label{tab:2dcf}}
\end{table}

\begin{figure}[ht]
\centering
\includegraphics[width=0.48\textwidth]{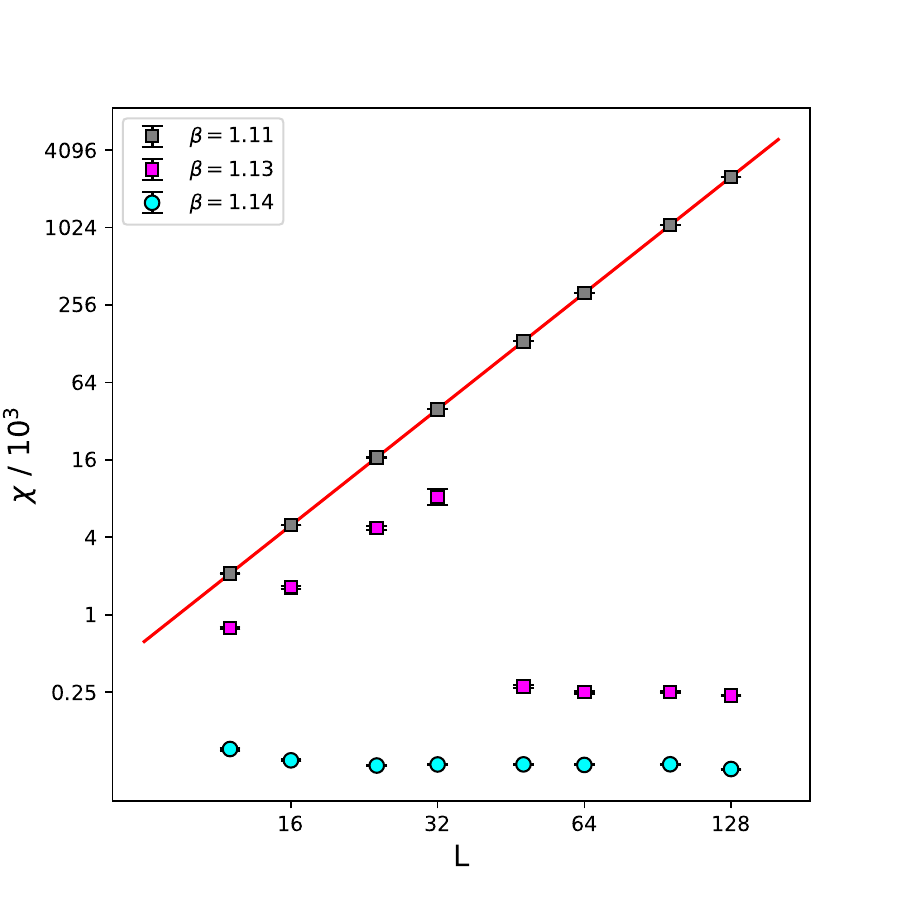}
\caption{Our \hbox{\ac{MCMC}} results for $\chi$ in $\dm=3$ in the $\SU(3)$ case as a function of $L$ at three different values of $\beta$. The deconfined phase is clearly visible at $\beta=1.11$ where the data fits to the form $\chi \sim A L^3$ for all our data from $L \geq 12$ and gives us $A=1.209(1)$. The confined phase is visible at $\beta=1.14$. The strange behavior of $\chi$ as a function of $L$ at $\beta=1.13$ is consistent with being in the vicinity of a first-order transition as explained in the text.
}
\label{fig:D3su3}
\end{figure}

Focusing on our results in $\dm=3$, we again find clear evidence for the confinement-deconfinement transition. In the case of $\SU(2)$, the transition seems to be continuous, in agreement with traditional lattice gauge theories \cite{Curci:1984yw,Das:1985up}. In \cref{fig:D3su2}, on the left, we plot $\chi$ as a function of $L$ for four different values of $\beta$ in the transition region. When $\beta=1.04$, we find a deconfined phase and $\chi$ grows as $A L^3$ where a fit gives $A=0.2400(7)$ with a $\chi^2/DOF \sim 1$ including data for $L \geq 48$. On the other hand, when $\beta=1.05$, we see $\chi$ saturates for large values of $L$. Assuming the transition to be second order, we expect $\chi$ to again obey the scaling relation \cref{eq:scal} in the three dimensional Ising universality class with $\eta=0.0366(8)$ and $\nu=0.6298(5)$ \cite{PhysRevB.59.11471}. Fixing these critical exponents to their central values and $\beta_c=1.04530$ obtained from a fit of $f(x)$ to a fourth-order polynomial, we plot the scaling function in \cref{fig:D3su2} on the right.

Finally we focus on $\SU(3)$ in $\dm=3$. In \cref{fig:D3su3}, we plot $\chi$ as a function of $L$ for three different values of $\beta$.
When $\beta=1.11$ our data fits to the form $A L^3$ with $A=1.209(1)$ for all our data $L \geq 12$ showing our model is in a deconfined phase. On the other hand when $\beta=1.14$, $\chi$ saturates quickly indicating that we are in a confined phase. The transition between the two phases in this case turns out to be first order. We can see this clearly at $\beta=1.13$ where $\chi$ initially increases for $L \leq 32$ but then falls to a much lower value and saturates for $L\geq 48$. This change occurs via large fluctuations in the Monte Carlo estimate of $\chi$ for $L \sim 32$ as expected for a system near a first-order transition. For small lattices, the system can tunnel to the deconfined phase and hence $\chi$, being proportional to $L^3$, is much larger than in the confined phase. Eventually, the tunneling is suppressed and the $\chi$ saturates to a much smaller value. The first-order transition is consistent with studies in traditional $\SU(3)$ lattice gauge theories \cite{Iwasaki:1991pc}.

\section{Quantum Lattice Gauge Theories}
\label{sec-6}

Our ultimate goal is to investigate quantum critical points in qubit-regularized gauge theories, where continuum quantum field theories emerge. To achieve this, we need to explore the physics of the coupling $\delta$ in \cref{eq:cHQmod}. While such an exploration is challenging in higher dimensions, it can be carried out more easily in $\dm=1$, as we demonstrate below.  

Consider a one-dimensional qubit-regularized $\SU(2)$ pure gauge theory, where the lattice geometry consists of a square plaquette chain with $L$ plaquettes. Such a plaquette chain has been studied previously using the more traditional \ac{KS} Hamiltonian \cite{Yao:2023pht}. To investigate how confinement is affected by the parameter $\delta$, we introduce heavy matter fields at two sites, $x=0$ and $x=\txtw$, and compute the corresponding ground state energy as a function of the separation distance $\txtw$ between them.

\begin{figure*}[t]
\centering
\includegraphics[width=0.8\textwidth]
{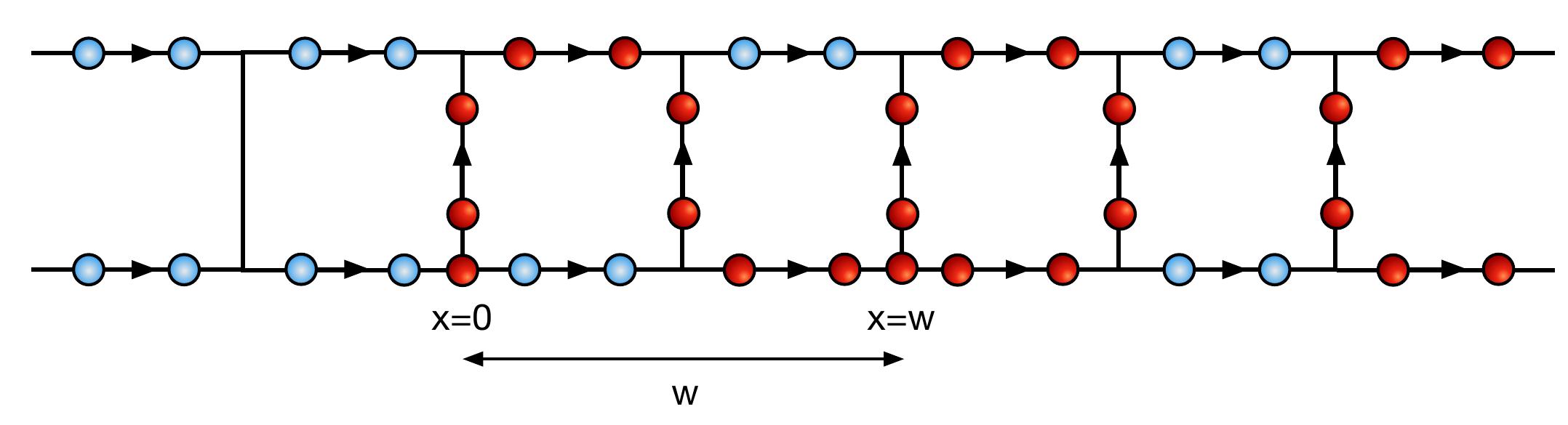}
\caption{A pictorial representation of a basis state in $\cH_\phys$ for plaquette chain with $\SU(2)$ gauge fields. Heavy matter fields have been introduced at sites on the lower rungs of $x=0$ and $x=\txtw$.
}
\label{fig:pchain}
\end{figure*}

The physical Hilbert space of the plaquette chain is straightforward to understand. We assign either a \(\mathbf{1}\) or \(\mathbf{2}\) irrep of $\SU(2)$ to each link. At sites where we introduce a heavy matter field, we use the representation \(\mathbf{2}\). Each assignment of irreps on links and sites corresponds to an \ac{MDTN} state on the chain, as illustrated in \cref{fig:pchain}. 

In the pure gauge theory, $\ldim = 1$ for all \ac{MDTN} states, allowing us to ignore the index $\alpha_s$. In the presence of matter, $\ldim$ can be either $1$ or $2$, depending on the dimer tensors on the links connected to the sites with matter. Nevertheless, as we will discuss below, choosing $c = 1$ for all plaquettes in the definition of \cref{eq:chupdef} ensures that one of the states decouples from the dynamics, allowing us, once again, to ignore the index $\alpha_s$.  

Note that our plaquette chain is equivalent to a ladder, and for convenience, we introduce heavy matter fields at sites $x=0$ and $x=\txtw$ on the lower chain as shown in \cref{fig:pchain}. It is easy to see that in the absence of matter fields, the \(\mathbf2\) links form closed loops, including those that may close around the boundary of the circle. In fact, the Hilbert space splits into a direct sum of two Hilbert spaces representing distinct topological classes of gauge field lines: $\cH_\phys^E$ (even) and $\cH_\phys^O$ (odd). Locally, we can distinguish these two sectors of the Hilbert space by simply looking at the link \acp{irrep} on the bottom and top links. Focusing on just these two links of a given plaquette $P$, there are four possible basis kets: \(\bket11\), \(\bket12\), \(\bket21\), and \(\bket22\), where the first label is the link \ac{irrep} on the top chain and the second label, on the bottom chain. We will refer to this as the local plaquette Hilbert space $\cH_P$ for later convenience. If there are no matter fields, neighboring plaquette states impose unique restrictions on the representations allowed on the rung. It is easy to argue that in the pure gauge theory, every plaquette can either be in the even sector with basis states \(\bket11\) and \(\bket22\), or in the odd sector with basis states \(\bket12\) and \(\bket21\).

A single site with a matter field in the fundamental representation induces a jump between the even and odd sectors on the two plaquettes on either side of it. In the discussion below, we will assume that $P_{x}$ labels the plaquette between sites $x$ and $x+1$. It is then easy to see that, in the illustration of \cref{fig:pchain}, the Hilbert spaces of plaquettes $P_{x}$ for $x < 0$ and $x \geq \txtw$ are in the even sector, while for $x \geq 0$ and $x < \txtw$, they are in the odd sector. We will refer to the physical Hilbert space of such a plaquette chain with matter fields on the lower rungs at $x = 0$ and $x = \txtw$ as $\cH_\phys^\txtw$, where $0 \leq \txtw \leq L$, with the definitions $\cH_\phys^{\txtw=0} \equiv \cH_\phys^E$ for the even sector and $\cH_\phys^{\txtw=L} \equiv \cH_\phys^O$ for the odd sector.

Coming to the discussion of the Hamiltonian, there are several possible choices, depending on \(c(\mdck,\{\alpha_s'\},P)\). In this work, we will assume
\begin{align}
c(\mdck,\{\alpha_s'\},P) = 1\,.
\end{align}
With this choice, it is easy to argue that, even when $\ldim \neq 1$, a unique local singlet projection always participates in the dynamics. The plaquette operator annihilates all other orthogonal states. Thus, when labeling the \ac{MDTN} states, we can assume that $\alpha_s = 1$ describes the state that participates in the dynamics and ignore the $\alpha_s > 1$ states, as those belong to a different super-selection sector that is not of interest to us. In the discussion below, we will focus on the Hilbert space with $\alpha_s = 1$, effectively assuming that $\ldim = 1$.

Note that the Hamiltonian of the plaquette chain is block-diagonal within each $\cH_\phys^\txtw$ space, whose basis states can be written as a product of Plaquette Hilbert spaces $\cH_P$. The Hamiltonian can be written more concisely using the diagonal plaquette operators
\begin{align}
\gamma^{(P)}_1 &\equiv \bop1111\,, \quad
\gamma^{(P)}_2 \equiv \bop2222 \,, \nonumber \\
\gamma^{(P)}_3 &\equiv \bop1212\,, \quad
\gamma^{(P)}_4 \equiv \bop2121 \,,
\end{align}
and the off-diagonal plaquette operators
\begin{align}
\sigma^{(P)}_{1,e} &\equiv (\bop2211 + \bop1122)\,, \nonumber \\
\sigma^{(P)}_{1,o} &\equiv (\bop2112 + \bop1221)\,.
\end{align}
For convenience, we also define
\begin{align}
\gamma^{(P)}_e &\equiv \gamma^{(P)}_1 + \gamma^{(P)}_2, \quad
\gamma^{(P)}_o \equiv \gamma^{(P)}_3 + \gamma^{(P)}_4.
\end{align}
Using these, we define four nearest-neighbor plaquette Hamiltonians:
\begin{align}
H^e_x &= \Big\{ 2 - \delta\ (\sigma^{(P_{x})}_e \gamma_e^{(P_{x+1})} + \gamma_e^{(P_{x})} \sigma^{(P_{x+1})}_e) \nonumber \\
& \quad \ - 2 \ \gamma_1^{(P_{x})} \gamma_1^{(P_{x+1})} \Big\}\,, \label{eq:plaqHe}
\end{align}
\begin{align}
H^o_x &= \Big\{ 2 - \delta\ (\sigma^{(P_{x})}_o \gamma_o^{(P_{x+1})} + \gamma_o^{(P_{x})} \sigma^{(P_{x+1})}_o) \nonumber \\
& \quad \ - \gamma_3^{(P_{x})} \gamma_3^{(P_{x+1})} - \gamma_4^{(P_{x})} \gamma_4^{(P_{x+1})} \Big\}\,, \label{eq:plaqHo}
\end{align}
\begin{align}
H^{eo}_x &= \Big\{ \frac{5}{2} - \delta\ (\sigma^{(P_{x})}_e \gamma_o^{(P_{x+1})}) - \delta\ (\gamma_e^{(P_{x})} \sigma^{(P_{x+1})}_o) \nonumber \\
& \quad \ - 2\ (\gamma_1^{(P_{x})} \gamma_3^{(P_{x+1})}) - (\gamma_e^{(P_{x})} \gamma_4^{(P_{x+1})}) \Big\}\,, \label{eq:plaqHeo}
\end{align}
\begin{align}
H^{oe}_x &= \Big\{ \frac{5}{2} - \delta\ (\gamma_o^{(P_{x})} \sigma^{(P_{x+1})}_e) - \delta\ (\sigma^{(P_{x})}_o \gamma_e^{(P_{x+1})}) \nonumber \\
& \quad \ - 2\ (\gamma_3^{(P_{x})} \gamma_1^{(P_{x+1})}) - (\gamma_4^{(P_{x})} \gamma_e^{(P_{x+1})}) \Big\}\,, \label{eq:plaqHoe}
\end{align}
which connect plaquettes $P_{x}$ and $P_{x+1}$. In terms of these operators, we can define the quantum Hamiltonians $H^{(\txtw)}_Q$ that act on each of the Hilbert spaces $\cH_\phys^\txtw$. In the even Hilbert space sector (i.e., $\cH_\phys^0$), the Hamiltonian is given by
\begin{align}
H^{(0)}_Q = \sum_x H^e_x\,,
\end{align}
in the odd Hilbert space sector (i.e., $\cH_\phys^L$), by
\begin{align}
H^{(L)}_Q = \sum_x H^o_x\,,
\end{align}
whereas, for every value of $\txtw$ in the range $0 < \txtw < L$, the Hamiltonian is given by
\begin{align}
H^{(\txtw)}_Q =  H^{eo}_{L-1} + H^{oe}_{\txtw-1} + \sum_{x=0}^{\txtw-2} H^o_x + \sum_{x=\txtw}^{L-2} H^e_x\,, \label{eq:Hchain}
\end{align}
as is easy to verify.

\begin{figure}[t]
\centering
\includegraphics[width=0.45\textwidth]
{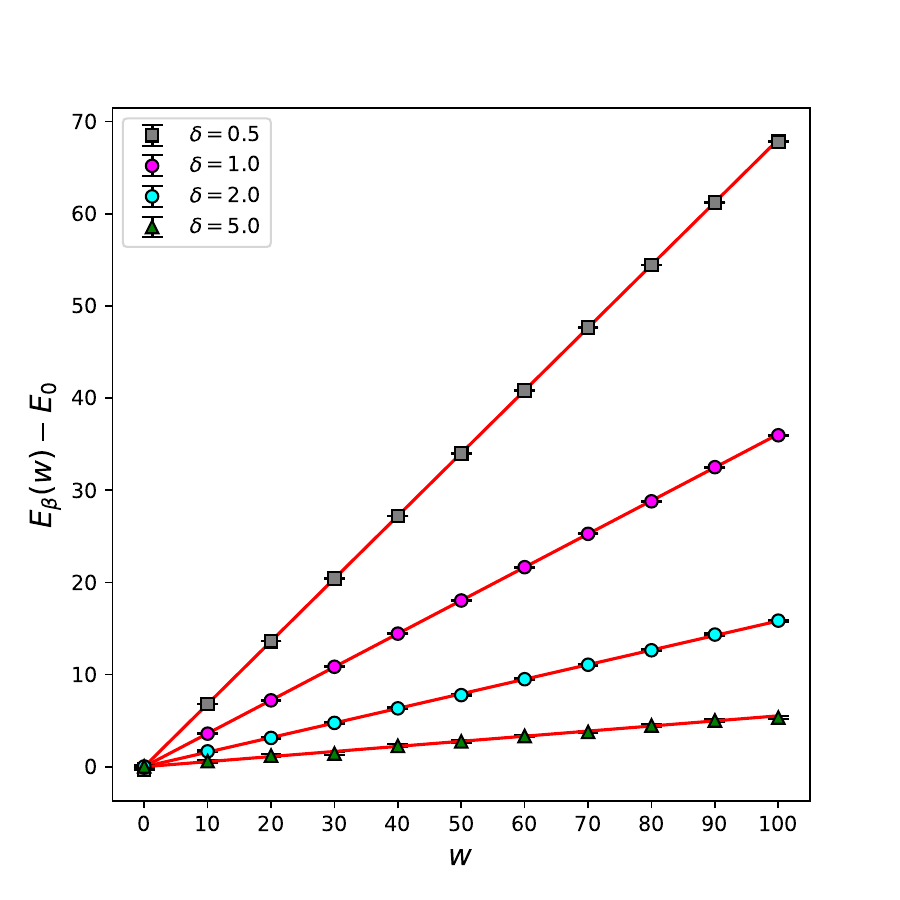}
\caption{Plot of $E_{\beta}(\txtw) - E_0$ as a function of $\txtw$, obtained for $L=100$ and $\beta=10$. The values of $E_0$ and the string tension $\sigma$ are given in \cref{tab:Edatfit}.}
\label{fig:Edat}
\end{figure}

In our current discussion, a natural observable to detect confinement is the average energy  
\begin{align}
E_{\beta}(\txtw) = \frac{\Tr_{\cH_\phys^\txtw} \Big(
H^{(\txtw)}_Q\ e^{-\beta H^{(\txtw)}_Q} \Big)}{\Tr_{\cH_\phys^\txtw} \Big(
e^{-\beta H^{(\txtw)}_Q} \Big)}\,.
\end{align}
At zero temperature, $E_{\beta}(\txtw)$ gives the ground state energy $E_g(\txtw)$, and in the confined phase, one expects  
\begin{align}
E_g(\txtw) = E_0 + \sigma \txtw 
\label{eq:conffit}
\end{align}
for large separations $\txtw$, where $\sigma$ is the string tension. In the deconfined phase, we expect $\sigma = 0$.  

We have implemented a Hamiltonian \hbox{\ac{MCMC}} method in continuous imaginary time to compute $E_{\beta}(\txtw)$ as a function of $\txtw$ for $L=100$ and $\beta=10$ at various values of $\delta \leq 5$. Increasing $\beta$ has only a small effect on the values of $E_{\beta}(\txtw)$, suggesting that $\beta=10$ effectively captures the physics of zero temperature.  

Our results are shown in \cref{fig:Edat}. The figure demonstrates that $E_{\beta}(\txtw)$ is consistent with the expected behavior of the ground state energy given in \cref{eq:conffit}, and we observe that $\sigma$ decreases as $\delta$ increases. The fitted values of $E_0$ and $\sigma$ are provided in \cref{tab:Edatfit}.

\begin{table}[ht]
\centering
\renewcommand{\arraystretch}{1.4}
\setlength{\tabcolsep}{4pt}
\begin{tabular}{c|c|c|c|c|c}
\TopRule
\multirow{2}{*}{$\delta$} & \multicolumn{3}{c|}{Monte Carlo} & 
\multicolumn{2}{c}{PT~\cref{eq:pt}} \\
\cline{2-6} 
 & $E_0$ & $\sigma$ & $\chi^2/DOF$ & $E_0$ & $\sigma$ \\
\MidRule
0.5 & -24.17(2) & 0.6805(3) & 0.34 & -55.75 & 1.000 \\
1.0 & -89.04(4) & 0.3608(6) & 0.40 & -90.50 & -0.375 \\
2.0 & -267.36(8) & 0.158(2) & 0.40 & -267.16 & 0.156 \\
5.0 & -856.1(2) & 0.055(2) & 0.51 & -856.12 & 0.055 \\
\BotRule 
\end{tabular}
\caption{The parameters $E_0$ and $\sigma$ in \cref{eq:conffit} computed using two different methods at $L=100$ and $\beta=10$. The values in columns two and three are obtained by fitting the \hbox{\ac{MCMC}} data for $E_{\beta}(\txtw)$ as a function of $\txtw$ at $L=100$ and $\beta=10$. In these fits, we exclude the values of $E_{\beta}(\txtw)$ at $\txtw=0$ and $\txtw=L$, as these correspond to energy values in the pure-gauge sector. The values in columns five and six are computed using \cref{eq:pt}, derived from perturbation theory, as explained in \cref{app-3}. \label{tab:Edatfit}}
\end{table}

Using the fact that our Hamiltonian is exactly solvable when $\delta=\infty$ and that the ground state is non-degenerate, we can compute $E_0$ and $\sigma$ as a power series in $1/\delta$ using non-degenerate perturbation theory. The details of this calculation up to third order are provided in \cref{app-3}. We obtain  
\begin{align}
E_{0}&\approx -2L\delta+\frac{3L}{2}-\frac{9L}{32}\delta^{-1}-\frac{L-1}{8}\delta^{-2}\,,\\
\sigma &\approx \frac{1}{4}\delta^{-1}+\frac{1}{8}\delta^{-2}\,.
\label{eq:pt}
\end{align}
The values of $E_0$ and $\sigma$ for $L=100$ are listed in columns five and six of \cref{tab:Edatfit}.  

We observe that $\delta=\infty$ corresponds to a deconfined quantum critical point, as the string tension vanishes in this limit. At this critical point, the parameter $1/\delta$ acts as a dimensionful (relevant) parameter, which, when introduced, leads to a nonzero string tension. This is analogous to the role of gauge couplings in $\dm=1$ in traditional lattice gauge theories. A similar but slightly more intricate deconfined critical point was recently discovered even in the presence of dynamical matter fields in a $\mathbb{Z}_2$ gauge theory \cite{Frank:2019jzv}.

\section{Conclusions}
\label{sec-7}

In this work, we have argued that an orthonormal basis for the physical Hilbert space of traditional lattice gauge theories can be constructed using the  irreps of the gauge group. Each basis state can be given a pictorial representation as a network of monomer and dimer tensors. This \ac{MDTN} basis state can be represented as $\ket{\mdck}$, where the set $\{\lambda_\ell\}$ labels the dimer-tensors, $\{\lambda_s\}$ labels the monomer-tensors, and the index $\{\alpha_s\}$ labels the different orthonormal gauge singlet states that can arise on the sites.

We then argued that the \ac{MDTN} basis states can be used to construct new types of qubit regularized gauge theories that are free of sign problems, and can be explored using \hbox{\ac{MCMC}} methods in any dimension. We introduced one simple Hamiltonian in \cref{eq:cHQmod}.
To show the richness of the physics of this simple Hamiltonian, we studied $\SU(2)$ and $\SU(3)$ gauge theories within the simple \hbox{\ac{ASQR}} scheme  \cite{Liu:2021tef}. 

We first introduced and studied classical lattice gauge theories by setting the parameter $\delta=0$ in \cref{eq:cHQmod}, ensuring that the Hamiltonian involves only local commuting operators that are diagonal in the $\ket{\mdck}$ basis. We argued that these classical gauge theories can effectively capture the physics of quantum gauge theories at finite temperatures. Using \hbox{\ac{MCMC}} methods, we demonstrated that the universal physics of finite-temperature confinement-deconfinement phase transitions in qubit-regularized $\SU(2)$ and $\SU(3)$ gauge theories in $\dm=2$ and $\dm=3$ dimensions is the same as that observed in traditional lattice gauge theories.

We then introduced quantum fluctuations by allowing a nonzero $\delta$ and studied the full physics of \cref{eq:cHQmod}. However, in this case, we focused on the $\dm=1$ case and demonstrated that the string tension in an $\SU(2)$ plaquette chain can be tuned to zero by increasing $\delta$. This clearly shows that $\delta$ plays the role of the gauge coupling in traditional lattice gauge theories. This result strongly suggests that it will play a similar role even in higher dimensions. 

The ultimate challenge, of course, is to construct Hamiltonians in higher dimensions that contain quantum critical points where continuum Yang-Mills theory can emerge. We view the Hamiltonian \cref{eq:cHQmod}, which we proposed in this work, as a starting point for such explorations. It is, of course, entirely possible that the desired fixed points of Yang-Mills theories will only emerge when higher irreps are included. This is straightforward within the general framework of the \ac{MDTN} basis states and would only mildly increase the complexity. 

It is also possible that simply adding extra layers in the \ac{ASQR} scheme could already lead to good effective field theories, similar to what was discovered in \cite{Zhou:2021qpm} within the context of the $O(4)$ spin model in $\dm=1$. This is essentially the main idea behind D-theory. Finally, the \ac{MDTN} basis states introduced in this work appear similar to the graph-theory-based formulation of $\SU(N)$ gauge theories proposed recently in \cite{Burbano:2024uvn}.

\section*{Acknowledgements}

The conceptualization and methodology, supervision and administration, as well as the first draft of the manuscript, were due to S.C.. The development of software, validation, first analyses, and visualization were carried out jointly by S.C. and R.X.S.. All three authors were involved in the investigation, interpretation, and editing of the manuscript, and all agreed on the final draft. Funding for this project was acquired by S.C. and T.B..

We would like to thank Ribhu Kaul,  Hanqing Liu, and Uwe-Jens Wiese for helpful discussions. We acknowledge the use of AI assistance, specifically ChatGPT~\hbox{\cite{openai2025chatgpt}}, in refining the language and clarity of this manuscript and providing citation to itself, before our final rounds of manual review and revision by all authors. This work was carried out in part under the auspices of the National Nuclear Security Administration of the U.S. Department of Energy at Los Alamos National Laboratory under Contract No. 89233218CNA000001. T.B. was also supported by the Quantum Science Center, a National Quantum Science Initiative of the Department of Energy, managed by Oak Ridge National Laboratory and by the U.S. Department of Energy, Office of Science—--High Energy Physics Contract KA2401032 to Los Alamos National Laboratory. S.C. and R.X.S. were supported by a Duke subcontract of this latter grant and also supported in part by the U.S. Department of Energy, Office of Science, Nuclear Physics program under Award No. DE-FG02-05ER41368.

\bibliographystyle{apsrev4-2} 
\showtitleinbib
\bibliography{fixapsbib,QC}
\onecolumngrid\vspace*{2\baselineskip}\twocolumngrid


\appendix

\section{Lattice Geometries}
\label{app-1}

In this appendix, we explain the geometry of the lattices used in our study of the confinement-deconfinement transition in \cref{sec-5}. Since the complexity of $\ldim$ depends on the coordination number of the lattice, keeping it small simplifies the calculations. For this reason, we consider lattices with a coordination number of $\dm+1$, where $\dm$ is the spatial dimension of the lattice. The explicit construction of the lattice geometry is discussed below.

\begin{table}[ht]
\centering
\renewcommand{\arraystretch}{1.4}
\setlength{\tabcolsep}{4pt}
\begin{tabular}{c|c|c|c}
\TopRule
\multicolumn{2}{c|}{$\SU(2)$} & \multicolumn{2}{c}{$\SU(3)$} \\
\hline
$\Hsg$ & $\ldim$ & $\Hsg$ & $\ldim$ \\
\MidRule
\multicolumn{4}{c}{Honeycomb Lattice}\\
\MidRule
$\otimesall{1111}$ & 1 &
$\otimesall{1111}$ & 1 \\
$\otimesall{1122}$ & 1 &
$\otimesall{113{\bar3}}$ & 1 \\
$\otimesall{2222}$ & 2 &
$\otimesall{1333}$ & 1 \\
- & - &
$\otimesall{1{\bar3}{\bar3}{\bar3}}$ & 1 \\
- & - &
$\otimesall{33{\bar3}{\bar3}}$ & 2 \\
\MidRule
\multicolumn{4}{c}{Diamond Lattice}\\
\MidRule
$\otimesall{11111}$ & 1 &
$\otimesall{11111}$ & 1 \\
$\otimesall{11122}$ & 1 &
$\otimesall{1113{\bar3}}$ & 1 \\
$\otimesall{12222}$ & 2 &
$\otimesall{11333}$ & 1 \\
- & - &
$\otimesall{11{\bar3}{\bar3}{\bar3}}$ & 1 \\
- & - &
$\otimesall{133{\bar3}{\bar3}}$ & 2 \\
- & - &
$\otimesall{{\bar3}3333}$ & 3 \\
- & - &
$\otimesall{3{\bar3}{\bar3}{\bar3}{\bar3}}$ & 3 \\
\BotRule 
\end{tabular}
\caption{ The dimension of the local singlet space ${\cal D}(\cH_g^s)$ for various possible choices of $\cH_s^g$ within our \hbox{\ac{ASQR}} on the \(d=2\) honeycomb and \(d=3\) diamond lattices. Since \({\cal D}(\cH_g^s)\) is permutation invariant and depends only on the representation label, permutations of the \(\cH_s^g\) are not shown. For those \(\cH_g^s\) that cannot be obtained by these rules, \({\cal D}(\cH_g^s)=0\). \label{tab:dimfac}}
\end{table}

We construct our lattice by starting with two periodic lattices of length $L$ in each direction, which we refer to as A (even) and B (odd). We define the lattice sites of the A-sublattice using the relation  
\begin{align}
\bfn_A \ =\ n_1 \bfe_1 \ +\ n_2 \bfe_2 \ + \dots +  n_\dm \bfe_\dm\,,
\end{align}
where $n_i = 0,1,2,\dots,L-1$ and $\bfe_i$ ($i = 1,2,\dots,\dm$) are independent crystal lattice vectors in $\dm$ dimensions. We then define the B-sublattice as  
\begin{align}
\bfn_B = \bfn_A + \ba\,,
\end{align}
where $\ba$ is a constant shift vector for all lattice sites $\bfn_A$. The $\dm+1$ nearest neighbors of the A-sublattice site $\bfn_A$ are given by $\bfn_A+\ba$, $\bfn_A+\ba-\bfe_1$, $\bfn_A+\ba-\bfe_2$, …, $\bfn_A+\ba-\bfe_\dm$. Similarly, the $\dm+1$ nearest neighbors of the B-sublattice site $\bfn_B$ are given by $\bfn_B-\ba$, $\bfn_B-\ba+\bfe_1$, $\bfn_B-\ba+\bfe_2$, …, $\bfn_B-\ba+\bfe_\dm$.  

The exact choice of $\ba$, $\bfe_1, \bfe_2, \dots, \bfe_\dm$ is not essential for defining the gauge theory, but they can be selected to maximize the spatial symmetry of the system. In particular, with an appropriate choice of these vectors, it is easy to see that our lattice corresponds to the usual one-dimensional linear lattice for $\dm=1$, the honeycomb lattice for $\dm=2$, and the diamond lattice for $\dm=3$.

With our definition of the lattice, \hbox{\ac{MDTN}} configurations $[\dc]$ and $[\dcxy]$ are well-defined once all link and site representations are specified. Using this information, we can compute $\ldim$ at all sites. For example, in $\dm=2$, each lattice site connects to three neighboring sites, and $\Hsg$ is a direct product of three link and one site Hilbert spaces. Similarly, in $\dm=3$, it is a direct product of four link and one site Hilbert spaces.  

In the \ac{ASQR} scheme, each $\lambda_s$ and $\lambda_\ell$ can take only the values $1$ or $2$ for $\SU(2)$, and $1$, $3$, or $\bar{3}$ for $\SU(3)$. This constraint limits the possible values of \(\ldim\), and we summarize all the nonzero values in \cref{tab:dimfac}.

\begin{figure}[thb]
\centering
\includegraphics[width=0.45\textwidth]{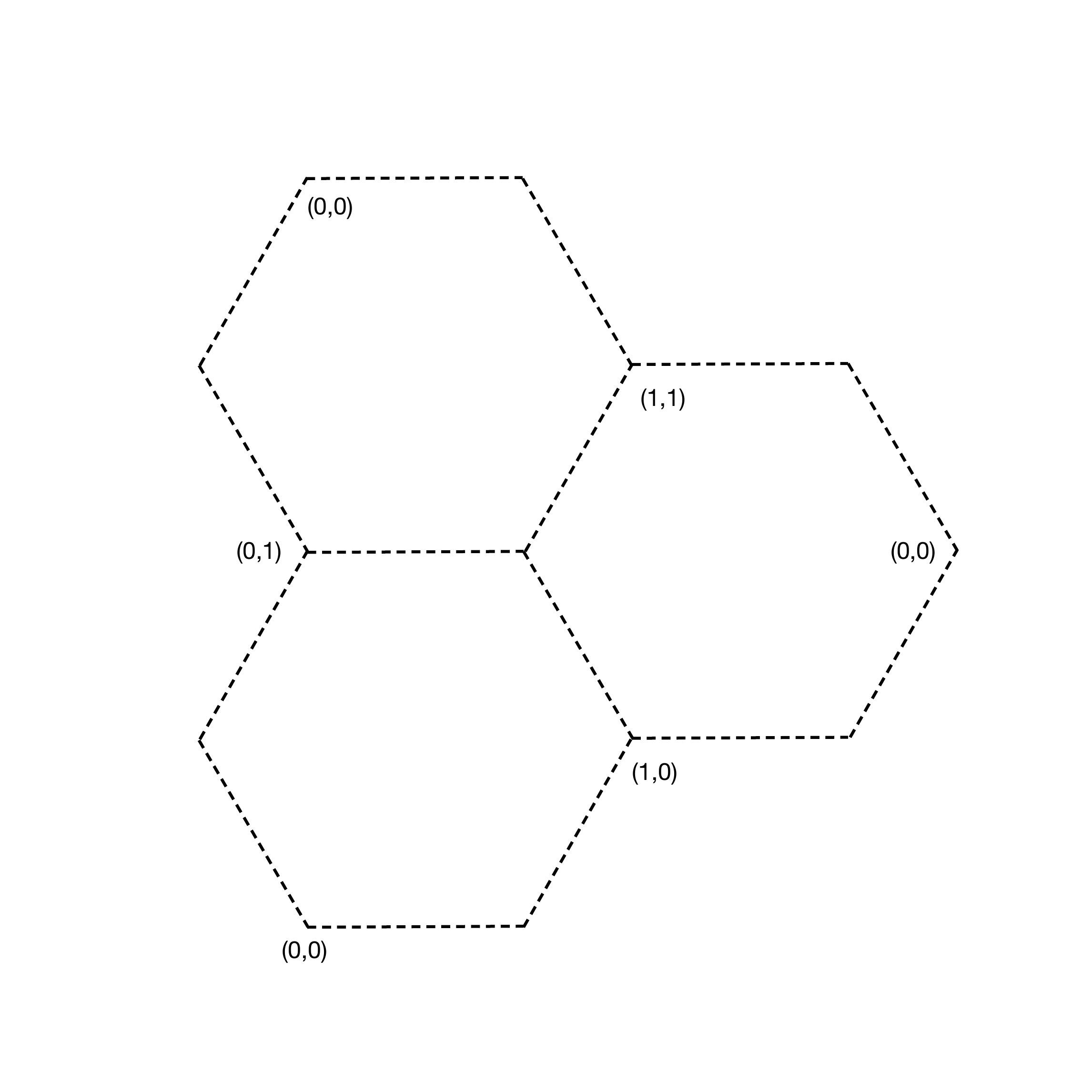}
\caption{The 8-site periodic lattice used for the comparison between exact and classical Monte Carlo results. The coordinates shown correspond to the $2\times 2$ $A$-sublattice. The tested algorithm was used to obtain the results presented in \cref{sec-5}.}
\label{fig:8-site}
\end{figure}

\section{Tests of the Monte Carlo Algorithms}
\label{app-2}

In this section, we provide evidence from small lattices that our \ac{MCMC} results agree with exact results. We test both our classical \ac{MCMC} algorithm, which was used to obtain results in \cref{sec-5}, and our quantum \ac{MCMC} algorithm, which was used to obtain results in \cref{sec-6}.

For the classical algorithms, we tested results on an 8-site lattice shown in \cref{fig:8-site} for both the $\SU(2)$ and $\SU(3)$ cases. We computed several observables but show results for the susceptibility $\chi$, as defined in \cref{eq:chi}, in \cref{tab:cmctest} for the $\SU(3)$ case.

\begin{table}[ht]
\centering
\renewcommand{\arraystretch}{1.4}
\setlength{\tabcolsep}{4pt}
\begin{tabular}{l|r|r}
\TopRule
\multirow{2}{*}{$\beta$} & \multicolumn{2}{c}{$\chi$} \\
\cline{2-3}
& exact & MC \\
\MidRule
0.0 & 12.6667 & 12.667(2) \\
0.2 & 11.3841 & 11.384(2) \\
0.4 & 9.9179 & 9.918(2) \\
0.6 & 7.9592 & 7.958(2) \\
0.8 & 5.5678 & 5.567(2) \\
1.0 & 3.5094 & 3.510(1) \\
\BotRule 
\end{tabular}
\caption{ \label{tab:cmctest} Comparision of exact versus \ac{MCMC} results for $\chi$ defined in \cref{eq:chi} on an $8$-site lattice shown in \cref{fig:8-site} for $\SU(3)$ gauge theory in the \ac{ASQR} scheme.}
\end{table}

For the quantum \ac{MCMC} we tested our results on a $L=12$ plaquette chain similar to the one shown in \cref{fig:pchain}. Our results are given in \cref{tab:qmctest}.
 
\begin{table}[ht]
\centering
\renewcommand{\arraystretch}{1.4}
\setlength{\tabcolsep}{4pt}
\begin{tabular}{c|c|c|c|c|c}
\TopRule
\multirow{2}{*}{$\txtw$} & \multicolumn{2}{c|}{$E_\beta(\txtw)$} & 
\multirow{2}{*}{$\txtw$} &
\multicolumn{2}{c}{$E_\beta(\txtw)$} \\
\cline{2-3} \cline{5-6}
& exact & MC & & exact & MC \\
\MidRule
0 & -0.10842 & -0.111(1) & 1 & 1.0291 &  1.026(1)\\
2 & 2.1704 & 2.170(1) & 3 & 3.2768 &  3.277(1)\\
4 & 4.3512 & 4.349(1) & 5 & 5.4052 &  5.405(1)\\
6 & 6.4478 & 6.446(1) & 7 & 7.4845 &  7.484(1) \\
8 & 8.5184 & 8.517(1) & 9 & 9.5516 & 9.551(1) \\
10 & 10.586 &  10.585(1) & 11 & 11.601 &  11.601(1) \\
12 & 12.263 & 12.262(1) & - & - & - \\
\BotRule 
\end{tabular}
\caption{ \label{tab:qmctest} Comparison of exact versus \ac{MCMC} results for an $12$-plaquette chain shown in \cref{fig:pchain}. These values are for $\delta=0.3$ and $\beta=1.0$. The energy $E_\beta(\txtw)$ was defined in \cref{sec-6}.}
\end{table}

\section{Large \texorpdfstring{$\delta$}{delta} Perturbation Theory}
\label{app-3}

In this appendix, we calculate the ground state energy $E_g$ of the chain Hamiltonian $H^{(\txtw)}_Q$ given in \cref{eq:Hchain} using large-$\delta$ perturbation theory. Since each plaquette $P_x$ uniquely belongs to either the even or the odd sector depending on $x$, we define  
$\sigma^{(P_x)}_1 = \sigma^{(P_x)}_{1,e}$ or  
$\sigma^{(P_x)}_1 = \sigma^{(P_x)}_{1,o}$ accordingly.  
Similarly, we define $\sigma^{(P_x)}_{3} = \gamma^{(P_x)}_2 - \gamma^{(P_x)}_1$ in the even sector and $\sigma^{(P_x)}_{3} = \gamma^{(P_x)}_4 - \gamma^{(P_x)}_3$ in the odd sector. With these definitions, we can express $H^{(\txtw)}_Q$ as  
\begin{align}
H^{(\txtw)}_Q \ =\ \delta \ H_0 + V\,,
\end{align}  
where  
\begin{align}
H_0 &= -2 \sum_x \sigma^{(P_x)}_1,
\end{align}  
and  
\begin{align}
V &= \frac{3}{2} L - \sum_{x} \left( \frac{1}{2} \sigma^{(P_x)}_{3} \sigma^{(P_{x+1})}_{3} -  
\theta(x) \sigma^{(P_x)}_{3} \right)\,,
\end{align}  
with $\theta(x) = 1$ for $\txtw \leq x < L$ and vanishing otherwise. In this notation, our system resembles the transverse-field Ising model with a uniform magnetic field in the region $\txtw \leq x < L$.

Since $H_0$ consists of decoupled terms, we can easily diagonalize it. To construct the eigenstates, let $\ket{\uparrow}$ and $\ket{\downarrow}$ be the normalized eigenkets of $\sigma^{(P)}_{3}$ with eigenvalues $+1$ and $-1$, respectively. The normalized eigenstates of $\sigma^{(P)}_{1}$ with eigenvalues $\pm 1$ are then given by  
\begin{align}
\ket{\pm 1} = \frac{\ket{\uparrow} \pm \ket{\downarrow}}{\sqrt{2}}\,.
\end{align}  
We can label the $2^L$ eigenstates of $H_0$ using the spin configuration $\{s\} = \{s_0, s_1, \dots, s_{L-1}\}$, where $\ket{s_x}$ (with $s_x = \pm 1$) represents the state of the plaquette $P_x$. For convenience, we uniquely label spin configurations using an index $k$, defined via the binary representation  
\begin{align}
k = \sum_{i=0}^{L-1} \frac{1}{2} (1 - s_i) 2^i\,,
\end{align}  
so that the corresponding eigenstates are  
\begin{align}
\ket{k} := \bigotimes_{i=0}^{L-1} \ket{s_i}\,,
\end{align}  
with eigenvalues  
\begin{align}
\varepsilon_k = -2 \sum_{i=0}^{L-1} s_i\,.
\end{align}  
Since the ground state of $H_0$ is non-degenerate, we compute the ground state energy of $H^{(\txtw)}_Q$ using non-degenerate Rayleigh-Schr\"odinger perturbation theory. Up to the third order, we have  
\begin{align}
E_g \approx \delta E^{(0)}_g + E^{(1)}_g + \frac{E^{(2)}_g}{\delta} + \frac{E^{(3)}_g}{\delta^2}\,,
\end{align}  
where $E^{(0)}_g = \varepsilon_0 = -2L$ is the ground state energy of $H_0$, obtained by setting $s_i = +1$ for all $i$. The first-order correction is  
\begin{align}
E^{(1)}_g = \langle 0 |V|0\rangle = \frac{3}{2} L\,.
\end{align}  
The second-order correction is  
\begin{align}
E^{(2)}_g &= \sum_{k\neq 0} \frac{|\bra{k} V \ket{0}|^2}{\varepsilon_{0} - \varepsilon_{k}}  
= -\frac{9L}{32} + \frac{\txtw}{4}\,,
\end{align}  
where the final result follows from choosing $\ket{k}$ such that only one or two consecutive spins are $-1$. The third-order correction is  
\begin{align}
E^{(3)}_g &= \sum_{k\neq 0} \sum_{k' \neq 0} \frac{\bra{0}V\ket{k} \bra{k}V\ket{k'} \bra{k'}V\ket{0}}{(\varepsilon_0-\varepsilon_k)(\varepsilon_0-\varepsilon_{k'})}
\nonumber \\
&\quad -\bra{0}V\ket{0} \sum_{k\neq 0} \frac{|\bra{k}V\ket{0}|^2}{(\varepsilon_0-\varepsilon_k)^2}\,,
\end{align}
which requires some effort to calculate, but the final result is simple and is given by
\begin{align}
E^{(3)}_g\ =\ -\frac{L-1}{8} + \frac{\txtw}{8}.
\end{align}  
Combining all these results, we obtain
\begin{align}
E_g \approx -2L \delta + \frac{3L}{2} -  
\left( \frac{9L}{32} - \frac{\txtw}{4} \right) \frac{1}{\delta}  
- \left( \frac{L - 1 - \txtw}{8} \right) \frac{1}{\delta^2}\,.
\end{align}  
By writing $E_g = E_0 + \sigma \txtw$, we identify the expressions for $E_0$ and $\sigma$ given in \cref{eq:pt}.

\end{document}